\documentclass[prb,twocolumn,groupedaddress,shownopacs,amssymb, amsmath]{revtex4-2}
\usepackage{graphicx,psfrag,amsmath,amssymb}

\usepackage[usenames, dvipsnames]{color}

\usepackage{hyperref}
\hypersetup{
    colorlinks=true,
    linkcolor=Blue,
    citecolor=Blue,
    filecolor=Blue,
    urlcolor=Blue,
    pdftitle={pair quantum geometry},
    }
\begin{document}

\title{Superconductivity beyond band geometry: emergence of pair quantum geometry}
\author{M. A. Keskiner and M. Iskin}
\affiliation{
Department of Physics, Ko\c{c} University, Rumelifeneri Yolu,
34450 Sar\i yer, Istanbul, T\"urkiye
}

\date{\today}

\begin{abstract}

The quantum geometry of Bloch states is known to shape the effective mass
of single particles in multiband systems. We show that this connection
extends to paired states. For a multiband Hubbard model with on-site
attraction, we derive an exact effective-mass theorem for two-body bound
states in vacuum and its counterpart for Cooper pairs near the critical
temperature within Gaussian fluctuation theory. In both cases, the inverse
mass contains, besides a ``conventional'' term from band dispersion and
band geometry, a distinct term governed by a quantum metric on the
manifold of paired states. This pair-geometric term requires a sublattice
pairing texture that varies with the pair momentum and, like the band
quantum metric, depends on the orbital embedding, whereas the total mass
does not. The one-body, two-body, and Cooper-pair theorems follow from a
single construction, with the band mass recovered as a special case. When
a collective mode lies inside the two-particle continuum, its continuation
to the second Riemann sheet renders the fluctuation kernel non-Hermitian,
leading to a biorthogonal quantum geometry and a complex Cooper-pair mass
whose real and imaginary parts describe propagation and damping. Exact
calculations for the sawtooth, Su-Schrieffer-Heeger, Hofstadter, and
fluorite-like lattices show that pair geometry contributes significantly
to the mass of bound pairs, and the fluorite-like lattice exhibits both an
undamped Cooper pair with a real mass and a damped one with a complex
mass. Pair quantum geometry thus emerges as an essential ingredient of
multiband superconductivity beyond band geometry.

\end{abstract}
\maketitle

\section{Introduction}
\label{sec:intro}

The internal geometry of Bloch states in momentum space, encoded in the
quantum geometric tensor of Provost and Vallee~\cite{Provost80}, has
emerged as a central organizing principle in condensed-matter physics.
Its imaginary part, the Berry curvature, underlies topological invariants
and quantized transport~\cite{Hasan10,qi11}, while its real part, the
quantum metric, which quantifies the gauge-invariant distance between
neighboring Bloch states in the Brillouin zone, has recently been
recognized as an equally fundamental quantity governing a broad range of
phenomena~\cite{resta11,torma23}. Together, they characterize
Bloch states in a way that the energy spectrum alone cannot.
Quantum geometry has become particularly important for
superconductivity, especially when conventional kinetic mechanisms are
strongly suppressed~\cite{torma22, peotta23, yu25, liu25, gao25, jiang25,
verma26, kitamura26}. The quantum metric of Bloch states contributes
directly to the superfluid weight~\cite{peotta15, liang17}, the
Cooper-pair effective mass~\cite{iskin18b, iskin24, huhtinen22, gao26},
and various pairing length scales~\cite{iskin23, chen23, thumin24,
iskin24c, iskin25, li25, virtanen25, lee26, xiao25, oh25, elden26,
chen26, kalkan26pair}. In flat or nearly flat bands, these contributions 
can even dominate~\cite{tian23}. These developments link the geometry
of the normal-state electronic structure to the properties of
the superconducting state~\cite{hu24, simon25, han24, dunbrack26, zhang26,
buthenhoff26}.

Yet superconductivity is fundamentally a problem of pairing rather than
single-particle motion~\cite{leggett,annett04}. The relevant low-energy
degrees of freedom are two-body bound states and many-body Cooper pairs,
which inhabit manifolds geometrically distinct from the single-particle
Bloch manifold. Paired states may therefore carry an intrinsic
quantum geometry that cannot be reduced to that of the constituent Bloch
states. Whether such a geometry leaves measurable signatures remains
largely unexplored~\cite{porlles23, guan26}. Despite the growing interest
in geometric aspects of superconductivity~\cite{torma22, peotta23, yu25,
liu25, gao25, jiang25, verma26, kitamura26}, existing approaches are cast
entirely in terms of Bloch-state geometry, and a framework for the
intrinsic geometry of the pairing manifold has not been established.

In this work, we develop such a framework. Starting from a multiband
Hubbard model with an on-site attractive interaction, we derive an exact
effective-mass theorem for two-body bound states in vacuum and its
many-body counterpart for Cooper pairs near the critical temperature
$T_c$ within Gaussian fluctuation theory. In both cases, the inverse
effective-mass tensor separates into two contributions. The first is a
``conventional'' term arising from the momentum curvature of the pair
kernel, which contains the band dispersion and the known
band-geometric effects~\cite{iskin23, iskin24}. The second is a new
geometric contribution governed by the momentum dependence of
the eigenvectors of the pair kernel, which we identify as the
\emph{pair quantum geometry}. This contribution is intrinsically
multiband: it vanishes identically in the single-band Hubbard model,
where the pairing amplitude is a scalar with no orthogonal eigenvector
subspace on which to define a metric, in contrast to the ``conventional''
term, which remains finite already at the single-band level.
The one-body effective-mass theorem is itself recovered as a
special case of the same construction.

Near $T_c$, an effective Ginzburg-Landau theory for the critical mode
can be obtained by integrating out the massive modes. This procedure
provides an alternative derivation of our Cooper-pair effective-mass
theorem and shows that the pair-geometric contribution to the gradient
coefficient originates from the momentum-dependent admixture of the
massive modes to the critical mode. Consequently, pair quantum geometry
enters the superconducting observables governed by the
Ginzburg-Landau gradient coefficients, including the coherence length,
penetration depth, superfluid weight, and upper critical field.

The resulting pair geometry is defined directly on the manifold of
paired states, distinct from the Bloch-state geometry, and
therefore represents a new level in the hierarchy of geometric
structures relevant to superconductivity. It vanishes when
the internal pairing texture is independent of the pair
momentum up to an overall phase, as for uniform pairing across
sublattices, or when the pair kernel becomes inversion symmetric in the
long-wavelength limit. A momentum-dependent pairing texture is
thus necessary for a nontrivial pair geometry, but not
sufficient, as additional symmetry constraints may still eliminate the
geometric contribution. Like the band quantum metric, however,
the pair geometry is defined relative to the orbital embedding. Its
separation from the ``conventional'' term depends on the embedding,
whereas the total mass does not. We verify the exactness of
the two-body effective-mass theorem and the quantitative importance of
the pair-geometric contribution numerically for four lattice
models spanning one, two, and three dimensions: the sawtooth chain, the
Su-Schrieffer-Heeger chain, the Hofstadter lattice, and a fluorite-like
lattice. In the many-body setting, the correspondence between the
two-body pair kernel and the inverse fluctuation propagator near $T_c$
carries the same geometric structure over to
Cooper pairs. When the collective mode lies inside the
two-particle continuum, analytic continuation to the second
Riemann sheet renders the fluctuation kernel non-Hermitian, requiring a
biorthogonal extension of the formalism and leading to a complex
Cooper-pair effective-mass tensor whose real and imaginary parts describe
pair propagation and damping, respectively. For the
fluorite-like lattice, we compute this complex mass and the damped
dispersion of a Cooper pair inside the continuum and contrast it with an
undamped mode lying in a gap of the continuum.

The remainder of the paper is organized as follows. In
Sec.~\ref{sec:emt}, we present the effective-mass theorems for the
one-body, two-body, and Cooper-pair problems, together with
the analytic continuation of the fluctuation kernel and the
self-consistency equations for $T_c$ and $\mu$. In Sec.~\ref{sec:ni}, we
illustrate the theory using four lattice models, including a
damped Cooper pair inside the two-particle continuum.
Finally, Sec.~\ref{sec:conc} contains our
conclusions. Appendices~\ref{sec:appA}, \ref{sec:apppair}, and
\ref{sec:appB} provide an alternative derivation of the two-body
effective-mass theorem, discuss the conditions for a nontrivial pair
geometry, and present an alternative derivation of the Cooper-pair
effective-mass theorem, respectively.

\section{Effective-mass theorems}
\label{sec:emt}

It is a relatively straightforward task to show that the effective
band mass in the one-body problem of a multiband lattice is directly
governed by the quantum geometry of the underlying Bloch
states~\cite{iskin19b}.
In close analogy with this result, here we systematically demonstrate
that the effective masses emerging in the two-body problem in vacuum,
and consequently in the analogous Cooper problems, are determined
by the quantum geometry associated with two-body and many-body
paired states.

\subsection{Multiband Hubbard model}
\label{sec:hm}

For this purpose, we consider a tight-binding Hubbard model defined on
a generic lattice with a multi-orbital basis. The Hubbard Hamiltonian
$
\mathcal{H}=\sum_\sigma \mathcal{H}_\sigma+\mathcal{H}_{\uparrow\downarrow}
$
consists of noninteracting and interacting contributions.
The single-particle part
$
\mathcal{H}_\sigma
=
-\sum_{ii'SS'}
t_{iS;i'S'}^\sigma\,
c_{Si\sigma}^\dagger c_{S'i'\sigma}
$
describes hoppings on the lattice, where $c_{Si\sigma}^\dagger$
creates a spin-$\sigma$ fermion on sublattice site
$S$ in unit cell $i$, and
$t_{iS;i'S'}^\sigma$ is the hopping amplitude from site $S'$ in unit cell
$i'$ to site $S$ in unit cell $i$.
To express $\mathcal{H}_\sigma$ in momentum space, we introduce the Fourier
transformation
$
c_{Si\sigma}^\dagger
=
\frac{1}{\sqrt{N_c}}
\sum_{\mathbf{k}}
e^{-i\mathbf{k}\cdot\mathbf{r}_{iS}}
c_{S\mathbf{k}\sigma}^\dagger,
$
where $N_c$ is the number of unit cells,
$\mathbf{k}=(k_x,k_y,k_z)$ is the crystal momentum in the first Brillouin
zone (with $\hbar=1$), and $\mathbf{r}_{iS}$ denotes the position of site
$S$ in unit cell $i$. The momentum sum is normalized according to
$
\sum_{\mathbf{k}\in\mathrm{BZ}}1=N_c.
$
For a lattice with $N_b$ sublattice sites per unit cell, the total number
of sites is $N=N_bN_c$.
In the sublattice basis, the Hamiltonian becomes
\begin{align}
\mathcal{H}_\sigma
=
\sum_{SS'\mathbf{k}}
(h^{SS'}_{\mathbf{k} \sigma} - \mu \delta_{SS'})
c_{S\mathbf{k}\sigma}^\dagger
c_{S'\mathbf{k}\sigma}, \label{eq: singleparticle}
\end{align}
where $\mu$ is the chemical potential and  the matrix elements
$
h^{SS'}_{\mathbf{k} \sigma}
=
-\frac{1}{N_c}
\sum_{ii'}
t_{iS;i'S'}^\sigma
e^{-i\mathbf{k}\cdot
(\mathbf{r}_{iS}-\mathbf{r}_{i'S'})}
$
form the Bloch Hamiltonian matrix $\mathbf{h}_{\mathbf{k} \sigma}$.
The corresponding single-particle spectrum
is determined by the eigenvalue equation
\begin{align}
\sum_{S'}
h^{SS'}_{\mathbf{k} \sigma}
n_{S'\mathbf{k}\sigma}
=
\varepsilon_{n\mathbf{k}\sigma}
n_{S\mathbf{k}\sigma},
\end{align}
where $\varepsilon_{n\mathbf{k}\sigma}$ is the energy of the $n$th Bloch
band and $n_{S\mathbf{k}\sigma}$ is the periodic part of the corresponding
Bloch eigenstate.
The corresponding real-space eigenstates are
$
\langle i S | n \mathbf{k} \sigma \rangle =
\frac{e^{i \mathbf{k} \cdot \mathbf{r}_{i S}}}{\sqrt{N_c}}
n_{S\mathbf{k}\sigma}.
$
Using the transformation
$
c_{S\mathbf{k}\sigma}^\dagger
=
\sum_n
n_{S\mathbf{k}\sigma}^*
c_{n\mathbf{k}\sigma}^\dagger,
$
the noninteracting Hamiltonian is diagonalized in the band basis,
\begin{align}
\mathcal{H}_\sigma =
\sum_{n\mathbf{k}}
\xi_{n\mathbf{k}\sigma}
c_{n\mathbf{k}\sigma}^\dagger
c_{n\mathbf{k}\sigma}.
\end{align}
where
$
\xi_{n\mathbf{k}\sigma} = \varepsilon_{n\mathbf{k}\sigma} - \mu
$
is the shifted dispersion with respect to the chemical potential $\mu$.

The interaction term,
$
\mathcal{H}_{\uparrow\downarrow}
=
-U\sum_{iS}
c_{Si\uparrow}^\dagger
c_{Si\downarrow}^\dagger
c_{Si\downarrow}
c_{Si\uparrow},
$ with $U\ge0$, describes an on-site attractive interaction between
opposite-spin fermions occupying the same lattice site.
In momentum space, parameterized by the total center-of-mass momentum
$\mathbf{q}$, it can be written as
\begin{align}
\mathcal{H}_{\uparrow\downarrow}
=
-\frac{U}{N_c}
\sum_{S\mathbf{k}\mathbf{k}'\mathbf{q}}
c_{S \mathbf{k} \uparrow}^\dagger
c_{S,-\mathbf{k}+\mathbf{q},\downarrow}^\dagger
c_{S,-\mathbf{k}'+\mathbf{q},\downarrow}
c_{S \mathbf{k}' \uparrow}.
\end{align}
Although finite-range density-density interactions $U_{iS; i'S'}$ can
induce both spin-singlet and spin-triplet bound states in multi-orbital
systems~\cite{iskin24b}, we focus here for simplicity on this purely
local term; a systematic treatment of the remaining interaction channels
permitted by symmetry in a multiorbital setting, e.g., Hund exchange,
pair-hopping, and spin-exchange terms, is left for future work.

\subsection{One-body problem}
\label{sec:obp}

Next, we establish a direct connection between the band curvature and
band geometry for a single particle, which forms the foundation for
the analogous structure that emerges in the two-body problem in vacuum
and in the Cooper problem. The first derivative of the band energy
follows from the Hellmann-Feynman theorem,
$
{\partial_{k_i} \varepsilon_{n\mathbf{k}\sigma} =
\langle n_{\mathbf{k}\sigma} |
\partial_{k_i} \mathbf{h}_{\mathbf{k}\sigma}
| n_{\mathbf{k}\sigma} \rangle},
$
where we introduce the shorthand notation
$
{\partial_{k_i} \equiv \frac{\partial}{\partial k_i}}
$
for partial derivatives. Taking an additional derivative with respect to
$k_j$ yields
$
\partial^2_{k_i k_j} \varepsilon_{n\mathbf{k}\sigma} =
\langle \partial_{k_j} n_{\mathbf{k}\sigma} |
\partial_{k_i} \mathbf{h}_{\mathbf{k}\sigma} |
n_{\mathbf{k}\sigma} \rangle +
\langle n_{\mathbf{k}\sigma} | \partial^2_{k_i k_j} \mathbf{h}_{\mathbf{k}\sigma}
| n_{\mathbf{k}\sigma} \rangle +
\langle n_{\mathbf{k}\sigma} | \partial_{k_i} \mathbf{h}_{\mathbf{k}\sigma} |
\partial_{k_j} n_{\mathbf{k}\sigma} \rangle.
$
To evaluate the terms involving derivatives of the Bloch states, we
differentiate the eigenvalue equation with respect to $k_j$ and project
onto a distinct band $m \neq n$, leading to
$
{\langle m_{\mathbf{k}\sigma}| \partial_{k_j}\mathbf{h}_{\mathbf{k}\sigma}|
n_{\mathbf{k}\sigma}\rangle = \langle m_{\mathbf{k}\sigma}|
\partial_{k_j} n_{\mathbf{k}\sigma}\rangle
(\varepsilon_{n\mathbf{k}\sigma}-\varepsilon_{m\mathbf{k}\sigma})}.
$
Using this relation together with completeness in the subspace orthogonal
to $|n_{\mathbf{k}\sigma}\rangle$, the first term can be rewritten as
$
\langle \partial_{k_j} n_{\mathbf{k}\sigma}|
\partial_{k_i}\mathbf{h}_{\mathbf{k}\sigma}|
n_{\mathbf{k}\sigma}\rangle
= \sum_{m \neq n}
\langle \partial_{k_j} n_{\mathbf{k}\sigma}|
m_{\mathbf{k}\sigma}\rangle
\langle m_{\mathbf{k}\sigma}|
\partial_{k_i} n_{\mathbf{k}\sigma}\rangle
(\varepsilon_{n\mathbf{k}\sigma}-\varepsilon_{m\mathbf{k}\sigma}),
$
with an analogous expression for the third term by Hermitian conjugation.
Adding these contributions, the second derivative of the band energy defines
the matrix elements of the inverse effective-mass tensor
$\mathbf{M}_{n\mathbf{k}\sigma}^{-1}$ as~\cite{iskin19b}
\begin{equation}
\begin{split}
(M_{n\mathbf{k}\sigma}^{-1})_{ij}=
\langle n_{\mathbf{k}\sigma}|
\partial^2_{k_i k_j} \mathbf{h}_{\mathbf{k}\sigma}
|n_{\mathbf{k}\sigma}\rangle
+ \sum_{m\neq n}
(\varepsilon_{n\mathbf{k}\sigma}-\varepsilon_{m\mathbf{k}\sigma})
g_{n\mathbf{k}\sigma}^{mij}.
\label{eq: inverseMass}
\end{split}
\end{equation}
Here, $g_{n\mathbf{k}\sigma}^{mij}$ denotes the matrix elements of the
band-resolved quantum-metric tensor, defined as
\begin{equation}
g_{n\mathbf{k}\sigma}^{mij} = 2 \operatorname{Re}[
\langle \partial_{k_i} n_{\mathbf{k}\sigma} | m_{\mathbf{k}\sigma} \rangle
\langle m_{\mathbf{k}\sigma} | \partial_{k_j} n_{\mathbf{k}\sigma} \rangle],
\label{eq: bandresolved}
\end{equation}
where $\operatorname{Re}$ denotes the real part and $m \ne n$.
As the name suggests, summing over all bands $m \neq n$ yields the
quantum-metric tensor of band $n$ at momentum $\mathbf{k}$
for spin $\sigma$,
$
g_{n\mathbf{k}\sigma}^{ij} = \sum_{m \neq n}
g_{n\mathbf{k}\sigma}^{mij},
$
which quantifies the gauge-invariant distance between neighboring Bloch
states in the Brillouin zone. Since the eigenvalues are strictly real,
only the real parts of these interband matrix elements contribute to
the band curvature.

Equation~\eqref{eq: inverseMass} admits a transparent physical interpretation.
The first term is the conventional contribution arising from the direct
momentum curvature of the Bloch Hamiltonian, reflecting how the matrix
elements vary across the Brillouin zone. The second term is a purely
geometric interband contribution, where the band-resolved quantum metric
between target band $n$ and remote bands $m$ is weighted by the corresponding
energy separation. This explicit connection between band curvature
and band geometry provides the natural foundation for the
subsections that follow. In particular, it anticipates how a similar
hierarchical structure governs the effective mass of bound pairs in the
two-body problem, and ultimately the Cooper pair effective mass in the
many-body problem near $T_c$.

\subsection{Two-body problem in vacuum}
\label{sec:tbp}

We next turn to the two-body problem in vacuum to investigate
how the underlying band structure and its associated geometry are
reorganized in determining the effective mass of a bound
pair~\cite{iskin21}. Specifically, we set $\mu = 0$ in
Eq.~\eqref{eq: singleparticle} and restrict the analysis to a system
containing exactly two particles with conserved center-of-mass momentum
$\mathbf{q} = (q_x, q_y, q_z)$. Since the Hubbard interaction is purely
onsite, spin-triplet bound states are not allowed, and the most general
variational ansatz can be written as
\begin{equation}
|\Psi_{\mathbf{q}}\rangle = \sum_{nm\mathbf{k}} 
\alpha_{nm\mathbf{k}}^{\mathbf{q}} c_{n\mathbf{k}\uparrow}^{\dagger} 
c_{m,-\mathbf{k}+\mathbf{q},\downarrow}^{\dagger} |0\rangle.
\label{eq:ansatz}
\end{equation}
Here, $|0\rangle$ denotes the vacuum state, and
$
\alpha_{nm\mathbf{k}}^{\mathbf{q}} 
= \alpha_{mn,-\mathbf{k}+\mathbf{q}}^{\mathbf{q}}
$
are complex variational amplitudes for the spin-singlet bound 
states satisfying the normalization condition
$
\sum_{nm\mathbf{k}} 
|\alpha_{nm\mathbf{k}}^{\mathbf{q}}|^2 = 1.
$

The two-body energies $E_\mathbf{q}$ are obtained by minimizing
$
\langle \Psi_{\mathbf{q}} | \mathcal{H} - E_\mathbf{q} | \Psi_{\mathbf{q}} \rangle
$
with respect to $\alpha_{nm\mathbf{k}}^{\mathbf{q}}$, where
$E_\mathbf{q}$ serves as a Lagrange multiplier enforcing the normalization
of the variational state. This procedure leads to~\cite{iskin21}
\begin{align}
\big( \varepsilon_{n\mathbf{k}\uparrow} + \varepsilon_{m,-\mathbf{k}
+\mathbf{q},\downarrow} &- E_\mathbf{q} \big)
\alpha_{nm\mathbf{k}}^{\mathbf{q}}
\nonumber \\
&= \frac{U}{N_c} \sum_{S} \beta_{S\mathbf{q}}
n^*_{S\mathbf{k}\uparrow}
m^*_{S,-\mathbf{k}+\mathbf{q},\downarrow},
\label{eq: twobody}
\end{align}
where we have introduced the dressed sublattice amplitudes
\begin{align}
\beta_{S\mathbf{q}} = \sum_{nm\mathbf{k}}
\alpha^{\mathbf{q}}_{nm\mathbf{k}}
n_{S\mathbf{k}\uparrow}
m_{S,-\mathbf{k}+\mathbf{q},\downarrow}.
\end{align}
These amplitudes play the role of pairing order parameters in the
two-body problem, and nontrivial solutions indicate the presence of
bound states. Although Eq.~\eqref{eq: twobody} admits both bound-state
and scattering-state solutions, eliminating
$\alpha_{nm\mathbf{k}}^{\mathbf{q}}$ in favor of $\beta_{S\mathbf{q}}$
reduces the problem to a nonlinear eigenvalue equation that directly
governs the bound states,
\begin{equation}
\sum_{S'} G^{SS'}_{\ell\mathbf{q}} \beta_{S'\mathbf{q}} = 0,
\label{eq:negq}
\end{equation}
in sublattice space, where the matrix elements of the kernel
$\mathbb{G}_{\ell\mathbf{q}}$ are
\begin{equation}
\begin{split}
G_{\ell\mathbf{q}}^{SS'} = \frac{\delta_{SS'}}{U}
-\frac{1}{N_c} \sum_{nm\mathbf{k}}
\frac{
n_{S\mathbf{k}\uparrow}
m_{S,-\mathbf{k}+\mathbf{q},\downarrow}
n_{S'\mathbf{k}\uparrow}^*
m_{S',-\mathbf{k}+\mathbf{q},\downarrow}^*
}{
\varepsilon_{n\mathbf{k}\uparrow}
+ \varepsilon_{m,-\mathbf{k}+\mathbf{q},\downarrow}
- E_{\ell\mathbf{q}}
}.
\label{eq:kernel}
\end{split}
\end{equation}
For a given $\mathbf{q}$, the index $\ell$ labels the distinct
bound-state branches, and $E_{\ell\mathbf{q}}$ denotes the corresponding
dispersion. The bound-state energies are determined self-consistently by
the condition
$
\det \mathbb{G}_{\ell\mathbf{q}} = 0.
$
A solution represents a stable bound state when
$E_{\ell\mathbf{q}}$ lies outside the two-particle scattering continuum.
For a fixed $\mathbf{q}$, the upper and lower continuum edges are given by
$
\max (
\varepsilon_{n\mathbf{k}\uparrow}
+ \varepsilon_{m,-\mathbf{k}+\mathbf{q},\downarrow}
)
$
and
$
\min (
\varepsilon_{n\mathbf{k}\uparrow}
+ \varepsilon_{m,-\mathbf{k}+\mathbf{q},\downarrow}
),
$
respectively. Physically, $\mathbb{G}_{\ell\mathbf{q}}$, viewed as a
function of $E$, plays the role of an inverse propagator for the bound
states: its zeros determine the bound-state energies, corresponding to
poles of the associated propagator.

Once the desired bound-state branch is identified, which we denote as 
$\ell = 1$ without loss of generality, its long-wavelength expansion 
yields the corresponding pair dispersion. 
In this paper, we assume that the branch of interest 
satisfies $E_{1\mathbf{q}} = E_{1,-\mathbf{q}}$ and attains a nondegenerate 
branch extremum at $\mathbf{q} = \mathbf{0}$. Under these conditions, 
the linear term vanishes at the extremum, 
and the leading-order pair dispersion is therefore purely quadratic,
\begin{equation}
E_{1\mathbf{q}} = E_0 + \frac{1}{2}\sum_{ij}
\left(M_{2b}^{-1}\right)_{ij} q_i q_j + \cdots,
\label{eq: boundstate} 
\end{equation}
where $E_0$ denotes the bound-state threshold at the branch extremum, 
and $\mathbf{M}_{2b}^{-1}$ is the inverse 
effective-mass tensor of the bound pair considered in this work. 
In the remainder of this section, the matrix $\mathbb{G}_{1\mathbf{q}}$ 
refers to the kernel $\mathbb{G}_{\ell\mathbf{q}}$ defined in 
Eq.~\eqref{eq:kernel}, evaluated self-consistently at the bound-state 
energy $E_{1\mathbf{q}}$.

$\mathbf{M}_{2b}^{-1}$ can be obtained rigorously by exploiting the 
spectral properties of the matrix $\mathbb{G}_{1\mathbf{q}}$ and its 
associated null space. To this end, we invoke Jacobi's formula, which 
relates the derivative of a determinant to the adjugate matrix 
$\operatorname{adj}(\mathbb{M})$:
\begin{align}
\partial_x \det\mathbb{M} = \operatorname{Tr}\left[
\operatorname{adj}(\mathbb{M}) \partial_x \mathbb{M} \right].
\label{eqn:jacobi}
\end{align}
Here, $x$ denotes an arbitrary continuous parameter, such as a momentum 
component $q_i$, and $\operatorname{Tr}$ denotes the trace. The adjugate 
matrix $\operatorname{adj}(\mathbb{M})$ is defined as the transpose of 
the cofactor matrix,
$
[\operatorname{adj}(\mathbb{M})]_{ij} = (-1)^{i+j} C_{ji},
$
where $C_{ji}$ is the $(j,i)$ cofactor, i.e., the signed determinant of 
the $(n-1)\times(n-1)$ submatrix obtained by deleting row $j$ and column 
$i$ from $\mathbb{M}$. The adjugate satisfies the exact algebraic identity
$
\mathbb{M}\operatorname{adj}(\mathbb{M}) 
= \operatorname{adj}(\mathbb{M})\mathbb{M}
= \det(\mathbb{M})\,\mathbb{I},
$
which holds for any square matrix $\mathbb{M}$, whether invertible or 
singular, where $\mathbb{I}$ denotes the identity matrix. 
Being polynomial in the matrix elements, the adjugate remains
well defined and smooth when $\det\mathbb{M} = 0$, so that Jacobi's
formula holds at the bound-state pole, where $\mathbb{G}_{1\mathbf{q}}$
is singular.

Since $\mathbb{G}_{1\mathbf{q}}$ is an $N_b \times N_b$ Hermitian matrix, 
it admits the spectral decomposition
\begin{align}
\mathbb{G}_{1\mathbf{q}} 
= \sum_{\zeta}\lambda_{\zeta\mathbf{q}}
|v_{\zeta\mathbf{q}}\rangle
\langle v_{\zeta\mathbf{q}}|,
\label{eq: G1q}
\end{align}
where $\lambda_{\zeta\mathbf{q}}$ are real eigenvalues and 
$|v_{\zeta\mathbf{q}}\rangle$ are orthonormal eigenvectors. 
Since we already assumed that the bound-state branch $E_{1\mathbf{q}}$ 
of interest is nondegenerate in the $\mathbf{q}\to\mathbf{0}$ limit, 
it follows that, at the bound-state pole defined by the 
secular equation
\begin{align}
\det\mathbb{G}_{1\mathbf{q}} = 0,
\label{eqn:secularG}
\end{align}
exactly one eigenvalue vanishes. We label this eigenvalue by $\zeta = 1$, 
i.e., the pole condition is 
\begin{align}
\lambda_{1\mathbf{q}} = 0,
\label{eqn:poleL}
\end{align}
with the associated null eigenvector $|v_{1\mathbf{q}}\rangle$ satisfying
$
\mathbb{G}_{1\mathbf{q}} |v_{1\mathbf{q}}\rangle = 0,
$
while all remaining eigenvalues satisfy 
$\lambda_{\zeta \mathbf{q}} \neq 0$ for $\zeta \neq 1$.
Note that only the $\zeta = 1$ eigenvector
$
|v_{1\mathbf{q}}\rangle \equiv 
(\beta_{A\mathbf{q}}, \beta_{B\mathbf{q}}, \ldots)^\mathrm{T}
$
is associated with the physical bound state $E_{1 \mathbf{q}}$, 
where $\mathrm{T}$ denotes transpose. The remaining $\zeta \ne 1$ 
eigenvectors are therefore referred to as auxiliary eigenvectors
of $\mathbb{G}_{1\mathbf{q}}$.
At the bound-state pole, the adjugate therefore reduces to the 
rank-one Hermitian operator
$
\operatorname{adj}(\mathbb{G}_{1\mathbf{q}})
=
\mathcal{C}_\mathbf{q}^{2b}\,
|v_{1\mathbf{q}}\rangle
\langle v_{1\mathbf{q}}|,
$
where
$
\mathcal{C}_\mathbf{q}^{2b} = \prod_{\zeta \neq 1}\lambda_{\zeta\mathbf{q}}
$
is real and nonzero, and the null eigenvector is normalized according to
$
\langle v_{1\mathbf{q}}|v_{1\mathbf{q}}\rangle = 1.
$

Equation~\eqref{eqn:secularG} holds identically along the bound-state branch
$E_{1\mathbf{q}}$ of interest. Differentiating it on shell with respect to
$q_i$, using Jacobi's formula Eq.~\eqref{eqn:jacobi} together with the
rank-one form of the adjugate, yields
$
\mathcal{C}_\mathbf{q}^{2b}\langle v_{1\mathbf{q}}|
\partial_{q_i}\mathbb{G}_{1\mathbf{q}}| v_{1\mathbf{q}}\rangle
+ \mathcal{C}_\mathbf{q}^{2b} \langle v_{1\mathbf{q}}|
\partial_{E_{1\mathbf{q}}}\mathbb{G}_{1\mathbf{q}}
|v_{1\mathbf{q}}\rangle
\partial_{q_i} E_{1\mathbf{q}} = 0.
$
At the branch extremum,
$
\partial_{q_i}E_{1\mathbf{0}} = 0,
$
and therefore
$
\langle v_{1\mathbf{0}}|
\partial_{q_i}\mathbb{G}_{1\mathbf{0}}|
v_{1\mathbf{0}}\rangle = 0
$
is the first-order consistency relation.
Differentiating Eq.~\eqref{eqn:secularG} once more with respect to 
$q_j$ on shell and evaluating the result at the extremum eliminates 
all terms proportional to
$
\partial_{q_j}E_{1\mathbf{0}} = 0.
$
The remaining contributions satisfy
$
\langle v_{1\mathbf{0}}|
\partial^2_{q_iq_j}\mathbb{G}_{1\mathbf{0}}
|v_{1\mathbf{0}}\rangle
+ \langle\partial_{q_j}v_{1\mathbf{0}}|
\partial_{q_i}\mathbb{G}_{1\mathbf{0}}
|v_{1\mathbf{0}}\rangle
+ \langle v_{1\mathbf{0}}|
\partial_{q_i}\mathbb{G}_{1\mathbf{0}}
|\partial_{q_j}v_{1\mathbf{0}}\rangle
+ \langle v_{1\mathbf{0}}
|\partial_{E_0}\mathbb{G}_{1\mathbf{0}}
|v_{1\mathbf{0}}\rangle
\partial^2_{q_iq_j} E_{1\mathbf{0}} = 0.
$
Using the identity
$
\partial_{q_i} \mathbb{G}_{1\mathbf{q}} | v_{1\mathbf{q}} \rangle
= -\mathbb{G}_{1\mathbf{q}} |\partial_{q_i} v_{1\mathbf{q}} \rangle,
$
together with its Hermitian conjugate, the terms involving derivatives of
the eigenvector reduce to
$
-2 \operatorname{Re} \langle \partial_{q_i} v_{1\mathbf{0}}
| \mathbb{G}_{1\mathbf{0}} | \partial_{q_j} v_{1\mathbf{0}} \rangle.
$
Rearranging the resulting expression for the inverse effective-mass tensor,
$
(M_{2b}^{-1})_{ij} \equiv
\partial^2_{q_j q_i} E_{1\mathbf{0}},
$
and substituting Eq.~\eqref{eq: G1q}, we obtain
\begin{equation}
(M_{2b}^{-1})_{ij} =
\frac{
\langle v_{1\mathbf{0}} |
\partial^2_{q_i q_j} \mathbb{G}_{1\mathbf{0}} | v_{1\mathbf{0}} \rangle
- \sum_{\zeta\neq 1}
\lambda_{\zeta\mathbf{0}}
g^{\zeta ij}_{1 \mathbf{0},2b}
}{
\langle v_{1\mathbf{0}} |
\partial_{-E_0} \mathbb{G}_{1\mathbf{0}}
| v_{1\mathbf{0}} \rangle
}.
\label{eq: pairmass}
\end{equation}
The second term in the numerator is governed by the branch-resolved quantum
metric associated with the two-body branch $\ell=1$, defined as
\begin{equation}
g^{\zeta' ij}_{\zeta \mathbf{q}, 2b} =
2 \operatorname{Re}[
\langle\partial_{q_i}v_{\zeta\mathbf{q}}
|v_{\zeta'\mathbf{q}}\rangle
\langle v_{\zeta'\mathbf{q}}
|\partial_{q_j}v_{\zeta\mathbf{q}}\rangle],
\label{eq: pairboundresolvedmetric}
\end{equation}
for $\zeta\neq\zeta'$. This quantity is the direct analogue of
Eq.~\eqref{eq: bandresolved}. However, unlike the physical bound-state
branch indexed by $\zeta=1$, which corresponds to the $\ell=1$ branch
considered above, the index $\zeta'$ labels the remaining eigenvectors of
the kernel $\mathbb{G}_{1\mathbf{q}}$ defined in Eq.~\eqref{eq: G1q}.
Consequently, Eq.~\eqref{eq: pairboundresolvedmetric} is constructed
entirely from the spectral decomposition of $\mathbb{G}_{1\mathbf{q}}$
and does not explicitly involve the other physical bound-state branches
with $\ell\neq1$.
An alternative but equivalent derivation of Eq.~\eqref{eq: pairmass},
based on the on-shell differentiation of the secular equation
Eq.~\eqref{eqn:secularG} together with the pole condition
Eq.~\eqref{eqn:poleL}, rather than Jacobi's formula
Eq.~\eqref{eqn:jacobi}, is presented in App.~\ref{sec:appA}.

The inverse effective-mass tensor of Eq.~\eqref{eq: pairmass} depends
only on the eigenvalue branch $E_{1\mathbf{q}}$ and is therefore
invariant under any change of representation of the pair kernel,
$\mathbb{G}_{1\mathbf{q}} \to \mathcal{U}_{\mathbf{q}}^\dagger
\mathbb{G}_{1\mathbf{q}} \mathcal{U}_{\mathbf{q}}$ with unitary $\mathcal{U}_{\mathbf{q}}$.
Its two terms are not, since the transverse overlaps change as
$\langle v_{\zeta\mathbf{q}} | \partial_{q_i} v_{1\mathbf{q}} \rangle
\to \langle v_{\zeta\mathbf{q}} | \partial_{q_i} v_{1\mathbf{q}} \rangle
- i \langle v_{\zeta\mathbf{q}} | \mathcal{W}_i | v_{1\mathbf{q}} \rangle$,
with $\mathcal{W}_i = -i(\partial_{q_i} \mathcal{U}_{\mathbf{q}}) \mathcal{U}_{\mathbf{q}}^\dagger$,
which shifts weight between the two terms unless
$\langle v_{\zeta \mathbf{0}} | \mathcal{W}_i | v_{1\mathbf{0}} \rangle = 0$
for all $\zeta \neq 1$, e.g., for a $\mathbf{q}$-independent $\mathcal{U}$ or a
$\mathbf{q}$-dependent overall phase. The physically relevant
$\mathbf{q}$-dependent case is a change of orbital embedding,
$\mathbf{r}_{iS} \to \mathbf{r}_{iS} + \mathbf{d}_S$, which leaves the
real-space Hamiltonian unchanged but multiplies the pair amplitudes
$n_{S\mathbf{k}\uparrow} m_{S,-\mathbf{k}+\mathbf{q},\downarrow}$ by
$e^{-i\mathbf{q}\cdot\mathbf{d}_S}$ through the Fourier phases of the
Bloch states. This corresponds to
$(\mathcal{U}_{\mathbf{q}})_{SS'} = e^{i\mathbf{q}\cdot\mathbf{d}_S} \delta_{SS'}$,
so that $(\mathcal{W}_i)_{SS'} = d_{Si}\, \delta_{SS'}$ is the operator of
orbital displacements. The two terms are thus separately embedding
dependent while their sum is not, exactly as for the frozen-pairing and
relaxation contributions to the superfluid weight~\cite{kalkan2026}, and
as for the curvature and Bloch-metric terms of the one-body result,
Eq.~\eqref{eq: inverseMass}, where
$\mathbf{h}_{\mathbf{k}\sigma}\to \mathcal{U}^\dagger_{\mathbf{k}}
\mathbf{h}_{\mathbf{k}\sigma}\mathcal{U}_{\mathbf{k}}$ leaves the band mass
invariant but redistributes weight between them, reflecting the known
embedding dependence of the Bloch quantum metric~\cite{huhtinen22, lee26}.
Since the on-site interaction fixes the sublattice basis, any other
$\mathbf{q}$-dependent change of basis would mix sublattices and replace
the on-site pair amplitudes $\beta_{S\mathbf{q}}$ by nonlocal
combinations, so that the decomposition is defined once the embedding
is fixed~\cite{kalkan2026}. In Sec.~\ref{sec:ni}, we adopt the embedding
in which each Bloch Hamiltonian is written. Eqs.~\eqref{eq: sawtooth}, \eqref{eq: ssh}, and \eqref{eq: hofstadter} place all orbitals of a unit cell at the same
point, $\mathbf{r}_{iS}=\mathbf{r}_{iA}$, which is the convention of
Refs.~\cite{iskin21, iskin24} with which the ``conventional'' term is
compared below, whereas Eq.~\eqref{eq: fluorite} places them at their
geometric positions, following Refs.~\cite{neves24, duan24}. The
decompositions in Figs.~\ref{fig:masspanel} and \ref{fig:masspanelsc}
refer to these embeddings.

The structure of the effective-mass theorem in Eq.~\eqref{eq: pairmass},
with $\lambda_{1\mathbf{q}} = 0$, closely mirrors its single-particle
counterpart in Eq.~\eqref{eq: inverseMass}. The first term in the numerator
corresponds to the so-called ``conventional'' contribution arising from the
direct momentum curvature of the kernel itself. We place ``conventional''
in quotation marks because this term generally contains contributions from
both the band dispersion and the band geometry. The second term, governed
by the resolvent sum over metric components, encodes the internal quantum
geometry of the two-body manifold.
Thus, this spectral sum provides the precise two-body 
analogue of the interband geometric term in Eq.~\eqref{eq: inverseMass}, 
establishing a formal correspondence in which the properly normalized kernel
$
\mathbb{G}_{1\mathbf{q}} /
\langle v_{1\mathbf{0}} |
\partial_{-E_0} \mathbb{G}_{1\mathbf{0}}
| v_{1\mathbf{0}} \rangle,
$
the null eigenvector $|v_{1\mathbf{q}}\rangle$, the orthogonal eigenvectors 
$|v_{\zeta \ne 1, \mathbf{q}}\rangle$, and the correspondingly normalized 
eigenvalues $0$ and
$
\lambda_{\zeta\mathbf{0}} /
\langle v_{1\mathbf{0}} |
\partial_{-E_0} \mathbb{G}_{1\mathbf{0}}
| v_{1\mathbf{0}} \rangle
$
play roles analogous to $\mathbf{h}_{\mathbf{k}\sigma}$, 
$|n_{\mathbf{k}\sigma}\rangle$, $|m_{\mathbf{k}\sigma}\rangle$, and 
$\varepsilon_{n\mathbf{k}\sigma}$ and $\varepsilon_{m\mathbf{k}\sigma}$, 
respectively. 
In this mapping, the geometric coupling quantifies how momentum-induced 
variations of the target eigenvector $\zeta = 1$ project onto the 
orthogonal subspace $\zeta \neq 1$, weighted by the corresponding eigenvalues 
$\lambda_{\zeta\mathbf{q}}$ of the kernel matrix $\mathbb{G}_{1\mathbf{q}}$.
We have further verified numerically that the Zak phase (in one dimension) 
and the Berry-curvature-derived Chern number (in two dimensions) 
obtained from the auxiliary eigenvectors of $\mathbb{G}_{1\mathbf{q}}$ 
reproduce the topology of the corresponding two-body bound-state 
branches. This supports the interpretation of $\mathbb{G}_{1\mathbf{q}}$ 
as an effective Hamiltonian for paired states; 
see also Refs.~\cite{herzog22, Iskin2024}.

This analogy is more than formal, since Eq.~\eqref{eq: inverseMass} is itself a
special case of Eq.~\eqref{eq: pairmass}. To see this, we introduce the
one-body kernel
\begin{equation}
\mathbb{S}_{\mathbf{k}}(E) = \mathbf{h}_{\mathbf{k}\sigma} - E\,\mathbb{I},
\end{equation}
which plays the role of $\mathbb{G}_{1\mathbf{q}}$ for a single particle.
Its eigenvectors are the Bloch states $|m_{\mathbf{k}\sigma}\rangle$,
which do not depend on $E$, and its eigenvalues are
$\rho_{m\mathbf{k}}(E) = \varepsilon_{m\mathbf{k}\sigma} - E$. The secular
equation $\det \mathbb{S}_{\mathbf{k}}(E) =
\prod_{m}(\varepsilon_{m\mathbf{k}\sigma}-E) = 0$ is satisfied by the band
energies, and selecting band $n$ through the pole condition
$\rho_{n\mathbf{k}}(E) = 0$ places the kernel on shell,
$E = \varepsilon_{n\mathbf{k}\sigma}$. Just as the eigenvalues
$\lambda_{\zeta \mathbf{0}}$ in Eq.~\eqref{eq: pairmass} are evaluated at
the bound-state energy $E_{0}$, the remaining one-body eigenvalues are
evaluated at this energy,
$\rho_{m\mathbf{k}} \equiv \rho_{m\mathbf{k}}(\varepsilon_{n\mathbf{k}\sigma})
= \varepsilon_{m\mathbf{k}\sigma} - \varepsilon_{n\mathbf{k}\sigma}$ for
$m \neq n$, i.e., they are minus the band separations. Using, in addition,
$\partial^{2}_{k_i k_j}\mathbb{S}_{\mathbf{k}} =
\partial^{2}_{k_i k_j} \mathbf{h}_{\mathbf{k}\sigma}$ at fixed $E$ and
$\partial_{-E}\mathbb{S}_{\mathbf{k}} = \mathbb{I}$, so that the
denominator is unity, Eq.~\eqref{eq: pairmass} with
$\mathbf{q} \to \mathbf{k}$ and $\zeta \to m$ becomes
$(M^{-1}_{n\mathbf{k}\sigma})_{ij} =
\langle n_{\mathbf{k}\sigma}|\partial^{2}_{k_i k_j}
\mathbf{h}_{\mathbf{k}\sigma}|n_{\mathbf{k}\sigma}\rangle
- \sum_{m\neq n} \rho_{m\mathbf{k}}\, g^{mij}_{n\mathbf{k}\sigma}$,
which is precisely Eq.~\eqref{eq: inverseMass}. Although
Eq.~\eqref{eq: pairmass} was derived at a branch extremum, where
$\partial_{q_i}E_{1\mathbf{q}}=0$ eliminates the terms that mix the
momentum and energy dependences of the kernel, no such assumption is
needed here. Since $\rho_{n\mathbf{k}}(E)$ depends on $E$ only through
the subtracted energy, $\partial_E\rho_{n\mathbf{k}}=-1$ and
$\partial^2_E\rho_{n\mathbf{k}}=\partial_E\partial_{k_i}\rho_{n\mathbf{k}}=0$
at every $\mathbf{k}$, so that differentiating the on-shell condition
$\rho_{n\mathbf{k}}(\varepsilon_{n\mathbf{k}\sigma})=0$ twice along the
band gives $\partial^2_{k_ik_j}\varepsilon_{n\mathbf{k}\sigma}
=\partial^2_{k_ik_j}\rho_{n\mathbf{k}}\big|_{E=\varepsilon_{n\mathbf{k}\sigma}}$
directly, and Eq.~\eqref{eq: inverseMass} holds at every $\mathbf{k}$.
The one-body problem is thus the special case of a kernel that is linear
in energy, with $E$-independent eigenvectors and a unit denominator. For
$\mathbb{G}_{1\mathbf{q}}$, and for the many-body kernel
$\mathbb{K}_{1\mathbf{q}}$ introduced below, the dependence on the energy
argument, evaluated on shell at $E_{1\mathbf{q}}$ (or at the mode
dispersion $z_{1\mathbf{q}}$), is nonlinear, and the denominator supplies
the normalization of the pair. The one-body, two-body, and Cooper-pair
theorems are therefore special cases of a single construction based on
the isolated zero eigenvalue of a kernel.

It is worth emphasizing that Eq.~\eqref{eq: pairmass} is exact for the
two-body problem in vacuum, as demonstrated for several lattice models
in Sec.~\ref{sec:ni}, and that it reproduces previously established
results in appropriate limits. For instance, when the pairing
texture is independent of $\mathbf{q}$ up to an overall phase,
$|v_{1 \mathbf{q}} \rangle = e^{i \phi_\mathbf{q}} |v_1\rangle$, as under
the uniform-pairing condition
$|v_1\rangle = (1,1,\ldots, 1)^\mathrm{T}/\sqrt{N_b}$, the quantities
$g^{\zeta ij}_{1\mathbf{0}, 2b}$ vanish identically because
$|\partial_{q_i} v_{1\mathbf{q}}\rangle$ is parallel to
$|v_{1\mathbf{q}}\rangle$ and hence orthogonal to the auxiliary states
$|v_{\zeta \ne 1, \mathbf{q}} \rangle$. A nontrivial pair geometry
therefore requires a momentum-dependent sublattice texture of the
amplitudes $\beta_{S\mathbf{q}}$, which, as discussed in
App.~\ref{sec:apppair}, is necessary but not sufficient. Since a change
of embedding multiplies $\beta_{S\mathbf{q}}$ by
$e^{-i\mathbf{q}\cdot\mathbf{d}_S}$, this condition refers to the chosen
embedding. Furthermore, in the presence of both time-reversal symmetry
and uniform pairing, the $\mathbf{q = 0}$ limit of
$\partial^2_{q_i q_j}\mathbb{G}_{1\mathbf{0}}$ reduces to the
``conventional'' contributions that depend only on the band dispersions
and band geometry, thereby recovering exactly the corresponding results
reported in the recent literature~\cite{iskin21, iskin24}. 
Having established the exact two-body effective-mass theorem, 
we next extend our analysis to the Cooper problem near $T_c$ 
within the Gaussian-fluctuation framework of the BCS-BEC crossover.

\subsection{Cooper problem near the critical temperature}
\label{sec:Cp}

The one-body and two-body analyses of the preceding sections revealed 
a geometric hierarchy. Here, we show that this hierarchy extends 
to the many-body description of the superconducting state 
near the critical temperature $T_c$, and that the underlying mechanism 
originates from a structural correspondence between the 
inverse pair-fluctuation propagator $\boldsymbol{\Gamma}^{-1}(q)$ 
of the many-body effective action and the two-body kernel 
$\mathbb{G}_{1\mathbf{q}}$. 
Throughout, $q = (\mathbf{q}, i\nu_s)$ denotes a collective variable 
comprising the center-of-mass momentum $\mathbf{q}$ and the bosonic 
Matsubara frequency $\nu_s = 2\pi s T$, with 
$s = 0, \pm1, \pm2, \ldots$, where $T$ is the temperature in units 
such that $k_\mathrm{B} = 1$. 
The extension of the present formalism to the low-$\mathbf{q}$ 
collective modes at $T = 0$ is the subject of ongoing work.

For this purpose, we employ the Grassmann functional-integral 
formalism~\cite{sademelo93, iskin23}. Starting from the 
multiband Hubbard Hamiltonian, we first decouple the quartic 
interaction through a sublattice-dependent Hubbard-Stratonovich 
transformation by introducing the bosonic field 
$\Delta_S(q)$ as the sublattice-resolved pairing order parameter. 
The field is then decomposed as
$
\Delta_S(q) = \Delta_S \delta_{q0} + \Lambda_S(q),
$
where $\Delta_S$ is the uniform static saddle-point contribution and 
$\Lambda_S(q)$ represents the sublattice-resolved fluctuations. 
Since $\Delta_S \to 0$ near $T_c$, the pairing field
$
\Delta_S(q) \to \Lambda_S(q)
$
is entirely governed by fluctuations.
Integrating out the fermionic degrees of freedom then yields an effective 
bosonic action, which may be expanded in powers of the fluctuation fields as
$
\mathcal{S}_{\mathrm{eff}} = \mathcal{S}_0 + \mathcal{S}_2[\Lambda] + \cdots.
$
Here, $\mathcal{S}_0$ is the saddle-point contribution, and
$
\mathcal{S}_2 = \frac{N_c}{T} \sum_{SS'q}
\Lambda_S^*(q) \Gamma^{-1}_{SS'}(q) \Lambda_{S'}(q)
$
is the Gaussian contribution, where the matrix elements of the inverse 
fluctuation propagator are given by~\cite{iskin23}
\begin{align}
\Gamma^{-1}_{SS'}(q) = \frac{\delta_{SS'}}{U}
&+ \frac{1}{2N_c} \sum_{nm\mathbf{k}}\frac
{\mathcal{X}_{n\mathbf{k}\uparrow} + \mathcal{X}_{m,-\mathbf{k}+\mathbf{q},\downarrow}}
{i\nu_s - \xi_{n\mathbf{k}\uparrow} - \xi_{m,-\mathbf{k}+\mathbf{q},\downarrow}}
\nonumber \\
\times &
n_{S\mathbf{k}\uparrow} m_{S,-\mathbf{k}+\mathbf{q},\downarrow}
n^*_{S'\mathbf{k}\uparrow} m^*_{S',-\mathbf{k}+\mathbf{q},\downarrow}.
\label{eq:gammainv}
\end{align}
Here,
$
\mathcal{X}_{n\mathbf{k}\sigma} = \tanh[\xi_{n\mathbf{k}\sigma}/(2T)]
$
is the thermal factor.

The physical significance of the matrix $\boldsymbol{\Gamma}^{-1}(q)$ 
becomes apparent through its role in determining the dispersions of the 
collective modes $\omega_{\eta\mathbf{q}}$ near $T_c$~\cite{klimin21}. 

The collective modes are encoded in the analytic continuation of
Eq.~\eqref{eq:gammainv} from the Matsubara frequencies to the complex
plane, $i\nu_s \to z$. The resulting $\boldsymbol{\Gamma}^{-1}(\mathbf{q}, z)$
is a sum of simple poles located on the real axis at
$z = \xi_{n\mathbf{k}\uparrow} + \xi_{m,-\mathbf{k}+\mathbf{q},\downarrow}$,
i.e., at pair energies measured from $2\mu$. It is therefore analytic for
$\operatorname{Im} z \neq 0$ and satisfies
$\boldsymbol{\Gamma}^{-1}(\mathbf{q}, z^{*}) =
[\boldsymbol{\Gamma}^{-1}(\mathbf{q}, z)]^{\dagger}$.
In the thermodynamic limit, the poles with nonzero thermal weight 
$\mathcal{X}_{n\mathbf{k}\uparrow}+\mathcal{X}_{m,-\mathbf{k}+\mathbf{q},\downarrow}$ 
merge into branch cuts along the two-particle continua, whose union on the 
real axis we denote by $\mathcal{R}_{\mathbf{q}}$, except for pairs with a 
$\mathbf{k}$-independent energy, such as flat-band pairs, which remain
isolated poles. 

The collective modes are the poles of
$\boldsymbol{\Gamma}(\mathbf{q},z)=[\boldsymbol{\Gamma}^{-1}(\mathbf{q},z)]^{-1}$,
i.e., the roots of~\cite{Castin2020,klimin21}
\begin{subequations}\label{eq:modes}
\begin{align}
\det\boldsymbol{\Gamma}^{-1}(\mathbf{q},z_{\mathbf{q}})&=0,\qquad
z_{\mathbf{q}}\in\mathbb{C}\setminus\mathcal{R}_{\mathbf{q}},\label{eq:modesa}\\
\det\boldsymbol{\Gamma}^{-1}_{\downarrow}(\mathbf{q},z_{\mathbf{q}})&=0,\qquad
\mathrm{Im}\,z_{\mathbf{q}}\leq 0,\label{eq:modesb}
\end{align}
\end{subequations}
on the first (physical) Riemann sheet and on the second
Riemann sheet defined below, respectively. The first Riemann sheet,
$\mathbb{C}\setminus\mathcal{R}_{\mathbf{q}}$, contains both half planes, and
its boundary values on the real axis from above and below,
$\boldsymbol{\Gamma}(\mathbf{q},\omega\pm i0^{+})$, are the retarded and
advanced pair propagators, respectively.

The two conditions are connected through the values of
$\boldsymbol{\Gamma}^{-1}$ on the real axis. Taking $i\nu_s\to\omega\pm i0^+$ in
Eq.~\eqref{eq:gammainv} and using
$1/(x\pm i0^+)=\mathcal{P}(1/x)\mp i\pi\delta(x)$ gives
\begin{equation}
\boldsymbol{\Gamma}^{-1}(\mathbf{q},\omega\pm i0^+)
=\mathcal{A}(\mathbf{q},\omega)\mp i\mathcal{B}(\mathbf{q},\omega),
\label{eq:AB0}
\end{equation}
where the Hermitian matrix $\mathcal{A}$ is the principal-value part of
Eq.~\eqref{eq:gammainv}, and the spectral density
\begin{align}
\mathcal{B}_{SS'}&(\mathbf{q},\omega)=\frac{\pi}{2N_c}\sum_{nm\mathbf{k}}
(\mathcal{X}_{n\mathbf{k}\uparrow}+\mathcal{X}_{m,-\mathbf{k}+\mathbf{q},\downarrow})
n_{S\mathbf{k}\uparrow}m_{S,-\mathbf{k}+\mathbf{q},\downarrow} \nonumber \\
&\times
n^*_{S'\mathbf{k}\uparrow}m^*_{S',-\mathbf{k}+\mathbf{q},\downarrow}
\delta(\omega-\xi_{n\mathbf{k}\uparrow}
-\xi_{m,-\mathbf{k}+\mathbf{q},\downarrow})
\label{eq:AB}
\end{align}
is a Hermitian matrix that is nonzero only on $\mathcal{R}_{\mathbf{q}}$, with
$\omega\mathcal{B}\succeq0$ since
$\mathcal{X}_{n\mathbf{k}\uparrow}+\mathcal{X}_{m,-\mathbf{k}+\mathbf{q},\downarrow}$
has the sign of
$\xi_{n\mathbf{k}\uparrow}+\xi_{m,-\mathbf{k}+\mathbf{q},\downarrow}$.
In terms of $\mathcal{B}$, the continued kernel reads
$\boldsymbol{\Gamma}^{-1}(\mathbf{q},z)=\mathbb{I}/U+\int d\omega'\,
\mathcal{B}(\mathbf{q},\omega')/[\pi(z-\omega')]$.
Moreover, on the support of the $\delta$ function, the thermal weight is
$
\tanh[\xi_{n\mathbf{k}\uparrow}/(2T)]-\tanh[(\xi_{n\mathbf{k}\uparrow}-\omega)/(2T)]
=(\omega/2T)\,\mathrm{sech}^2[\xi_{n\mathbf{k}\uparrow}/(2T)]+\mathcal{O}(\omega^2),
$
so $\mathcal{B}(\mathbf{q},0)=0$ for every $\mathbf{q}$ and $\mathcal{B}$ grows
linearly in $\omega$ around $\omega=0$. Outside $\mathcal{R}_{\mathbf{q}}$,
the two boundary values coincide with $\mathcal{A}$, so the roots of
Eq.~\eqref{eq:modesa} there are the real roots of $\det\mathcal{A}=0$,
which are the undamped modes. For $T\geq T_c$, where
$\boldsymbol{\Gamma}^{-1}(\mathbf{q},0)\succeq0$ for all $\mathbf{q}$
(Sec.~\ref{sec: sc}), Eq.~\eqref{eq:modesa} has no complex
roots~\cite{thouless60}. For a putative root
$\boldsymbol{\Gamma}^{-1}(\mathbf{q},z)|v\rangle=0$ with $\mathrm{Im}\,z\neq0$
and $b(\omega')=\langle v|\mathcal{B}(\mathbf{q},\omega')|v\rangle$, the
imaginary and real parts of $\langle v|\boldsymbol{\Gamma}^{-1}(\mathbf{q},z)|v\rangle=0$
give $\int d\omega'\,b(\omega')/|z-\omega'|^2=0$ and
$1/U=\int d\omega'\,\omega' b(\omega')/[\pi|z-\omega'|^2]$, whence
$\langle v|\boldsymbol{\Gamma}^{-1}(\mathbf{q},0)|v\rangle
=-(|z|^2/\pi)\int d\omega'\,[b(\omega')/\omega']/|z-\omega'|^2<0$, a
contradiction. This is Thouless's argument, based on the sign of
$\tanh a+\tanh b$, in matrix form.

On $\mathcal{R}_{\mathbf{q}}$, $\boldsymbol{\Gamma}^{-1}$ jumps by
$2i\mathcal{B}$ across the real axis, and modes inside the continuum must be
sought by continuing $\boldsymbol{\Gamma}^{-1}$ through this cut, as for an
atom coupled to the radiation field~\cite{cohen92,Kurkjian2019}. Following
Nozi\`eres~\cite{nozieres1964,Kurkjian2019}, the continuation from the upper to
the lower half plane through an interval $\mathcal{I}$ of $\mathcal{R}_{\mathbf{q}}$ on
which $\mathcal{B}$ is analytic is
\begin{equation}
\boldsymbol{\Gamma}^{-1}_\downarrow(\mathbf{q},z)=
\begin{cases}
\boldsymbol{\Gamma}^{-1}(\mathbf{q},z), & \mathrm{Im}\,z>0,\\
\boldsymbol{\Gamma}^{-1}(\mathbf{q},z)-2i\mathcal{B}(\mathbf{q},z), & \mathrm{Im}\,z\leq0,
\end{cases}
\label{eq:cont}
\end{equation}
where $\mathcal{B}(\mathbf{q},z)$ is the analytic continuation of
$\mathcal{B}(\mathbf{q},\omega)$ from $\mathcal{I}$ into the lower half plane. The
continued kernel, and hence its roots, depends in general on the interval $\mathcal{I}$
through which the cut is crossed. 
On $\mathcal{I}$, $\boldsymbol{\Gamma}^{-1}_\downarrow(\mathbf{q},\omega-i0^+)=\mathcal{A}-i\mathcal{B}
=\boldsymbol{\Gamma}^{-1}(\mathbf{q},\omega+i0^+)$, i.e., the second Riemann sheet extends
the retarded kernel analytically below $\mathcal{I}$ and coincides with the first 
sheet above it. Its roots with $\mathrm{Re}\,z_{\mathbf{q}}\in \mathcal{I}$ are therefore
either real or damped.

A real root requires $(\mathcal{A}-i\mathcal{B})v=0$, and since $\mathcal{A}$ and
$\mathcal{B}$ are Hermitian and $\mathcal{B}$ is semidefinite at fixed $\omega$,
this holds only if $\mathcal{A}v=\mathcal{B}v=0$. Away from $\omega=0$, this
requires a mode decoupled from the continuum, whereas at $\omega=0$, where
$\mathcal{B}(\mathbf{q},0)=0$, it is satisfied by the critical mode at $T_c$.
Otherwise, $\det\boldsymbol{\Gamma}^{-1}_\downarrow$ has no zero on $\mathcal{I}$, and the
roots of Eq.~\eqref{eq:modesb} lie at $\mathrm{Im}\,z_{\mathbf{q}}<0$, where
$\mathcal{B}(\mathbf{q},z)$ is complex and
$\boldsymbol{\Gamma}^{-1}_\downarrow(\mathbf{q},z_{\mathbf{q}})$ is a
non-Hermitian matrix. Such a root describes a mode with a finite lifetime, set by
its decay into the two-particle continuum~\cite{Castin2020,klimin21,Kurkjian2019}.
We denote the roots of Eqs.~\eqref{eq:modes} by $z_{\eta\mathbf{q}}$, whose real
parts $\omega_{\eta\mathbf{q}}=\mathrm{Re}\,z_{\eta\mathbf{q}}$ give the mode
dispersions, and $-\mathrm{Im}\,z_{\eta\mathbf{q}}\geq0$ gives the damping rates.

A key observation linking the many-body and two-body sectors is that, 
under the formal replacements
$
i\nu_s +2\mu\to E_{\ell\mathbf{q}}
$
and
$
\mathcal{X}_{n\mathbf{k}\sigma} \to 1,
$
the inverse fluctuation propagator reduces exactly to the two-body kernel,
$
\boldsymbol{\Gamma}^{-1}(q) \to \mathbb{G}_{\ell\mathbf{q}}.
$
As a result, the two-body and many-body secular equations as well
as their pole conditions share the same algebraic structure.
In direct analogy with the two-body problem, we label the collective 
mode of interest by $\eta = 1$. Its long-wavelength behavior is 
characterized by the corresponding inverse effective-mass tensor. 
For this purpose, we assume that the mode satisfies
$
z_{1\mathbf{q}} = z_{1,-\mathbf{q}}
$
and is nondegenerate at $\mathbf{q} = \mathbf{0}$. 
Under these conditions, the linear gradient vanishes at
$\mathbf{q}=\mathbf{0}$, and the leading-order expansion of the 
dispersion is
\begin{equation}
z_{1\mathbf{q}} = z_0 +
\frac{1}{2}\sum_{ij}(M^{-1}_{\mathrm{Cp}})_{ij} q_i q_j + \cdots,
\label{eq: mode}
\end{equation}
where $z_0$ denotes the value of $z_{1\mathbf{q}}$ at
$\mathbf{q}=\mathbf{0}$, with $z_0=0$ at $T_c$ by the Thouless criterion
(Sec.~\ref{sec: sc}), and 
$\mathbf{M}^{-1}_{\mathrm{Cp}}$ is the corresponding inverse 
effective-mass tensor.
For notational convenience, we define
\begin{align}
\mathbb{K}_{1\mathbf{q}} =
\boldsymbol{\Gamma}^{-1}_{\downarrow}
(\mathbf{q}, z_{1\mathbf{q}}),
\end{align}
with $\boldsymbol{\Gamma}^{-1}_{\downarrow} \equiv
\boldsymbol{\Gamma}^{-1}$ when the mode lies outside
$\mathcal{R}_{\mathbf{q}}$, so that the collective-mode dispersion is
determined by the secular equation
\begin{align}
\det \mathbb{K}_{1\mathbf{q}} = 0.
\label{eqn:secularK}
\end{align}
Since $\boldsymbol{\Gamma}^{-1}_\downarrow(\mathbf{q},z)$ is
analytic in $z$, $\mathbb{K}_{1\mathbf{q}}$ depends analytically on
$z_{1\mathbf{q}}$, so Eq.~\eqref{eqn:secularK} can be differentiated on
shell even when $z_{1\mathbf{q}}$ is complex. For a mode outside the
two-particle continuum, $z_{1\mathbf{q}}=\omega_{1\mathbf{q}}$ is real and
$\mathbb{K}_{1\mathbf{q}}=\mathcal{A}(\mathbf{q},\omega_{1\mathbf{q}})$ is
Hermitian. For a mode inside it, $\mathrm{Re}\,z_{1\mathbf{q}}\in \mathcal{I}$ and
$\mathrm{Im}\,z_{1\mathbf{q}}<0$ except at $\mathbf{q}=\mathbf{0}$ at $T_c$,
where $z_{1\mathbf{0}}=0$. Away from this point the kernel is evaluated
off the real axis and is non-Hermitian. This stands in contrast to the
two-body kernel $\mathbb{G}_{1\mathbf{q}}$, which is Hermitian by
construction. As a consequence, whereas the two-body analysis yields a
real inverse effective-mass tensor $\mathbf{M}^{-1}_{2b}$, the
Cooper-pair inverse effective-mass tensor $\mathbf{M}^{-1}_{\mathrm{Cp}}$
is generally complex, as discussed below.

To evaluate $\mathbf{M}_\mathrm{Cp}^{-1}$ for the collective mode of 
interest, we extend the adjugate-matrix approach developed for the 
two-body problem to the non-Hermitian kernel $\mathbb{K}_{1\mathbf{q}}$. 
This extension requires a biorthogonal basis of right and left 
eigenvectors, which exists as long as $\mathbb{K}_{1\mathbf{q}}$
is diagonalizable (away from exceptional points), satisfying
$
\mathbb{K}_{1\mathbf{q}} |u^R_{\eta\mathbf{q}}\rangle =
\chi_{\eta\mathbf{q}} |u^R_{\eta\mathbf{q}}\rangle
$
and
$
\langle u^L_{\eta\mathbf{q}}| \mathbb{K}_{1\mathbf{q}} =
\chi_{\eta\mathbf{q}} \langle u^L_{\eta\mathbf{q}}|,
$
together with the biorthonormality condition
$
\langle u^L_{\eta\mathbf{q}}|u^R_{\eta'\mathbf{q}}\rangle
=
\delta_{\eta\eta'}.
$
The kernel therefore admits the spectral resolution
\begin{align}
\mathbb{K}_{1\mathbf{q}}
=
\sum_{\eta}
\chi_{\eta\mathbf{q}}
|u^R_{\eta\mathbf{q}}\rangle
\langle u^L_{\eta\mathbf{q}}|,
\label{eq:K1q}
\end{align}
where the eigenvalues $\chi_{\eta\mathbf{q}}$ are generally complex.
At a collective-mode pole, one eigenvalue must vanish. Without loss of 
generality, we identify this mode with $\eta = 1$, so that the pole 
condition becomes
\begin{align}
\chi_{1\mathbf{q}} = 0.
\label{eqn:poleC}
\end{align}
The corresponding right and left eigenvectors span the null spaces
$
\mathbb{K}_{1\mathbf{q}}|u^R_{1\mathbf{q}}\rangle = 0
$
and
$
\langle u^L_{1\mathbf{q}}|\mathbb{K}_{1\mathbf{q}} = 0,
$
while all remaining eigenvalues satisfy
$
\chi_{\eta\mathbf{q}} \neq 0
$
for $\eta \neq 1$.
Upon imposing the normalization
$
\langle u^L_{1\mathbf{q}}|u^R_{1\mathbf{q}}\rangle = 1,
$
the spectral decomposition immediately yields the adjugate matrix at the 
pole in the rank-one form
$
\operatorname{adj}(\mathbb{K}_{1\mathbf{q}})
=
\mathcal{C}_{\mathbf{q}}^{\mathrm{Cp}}
|u^R_{1\mathbf{q}}\rangle
\langle u^L_{1\mathbf{q}}|,
$
where
$
\mathcal{C}_{\mathbf{q}}^{\mathrm{Cp}}
=
\prod_{\eta\neq1}\chi_{\eta\mathbf{q}}
$
is the product of the remaining nonvanishing eigenvalues. Since
$\mathbb{K}_{1\mathbf{q}}$ is evaluated on shell, this rank-one form holds
for every $\mathbf{q}$ and may be differentiated along the dispersion.

The derivation of $\mathbf{M}_\mathrm{Cp}^{-1}$ proceeds through the
systematic on-shell differentiation of Eq.~\eqref{eqn:secularK} with
respect to momentum. Differentiating with respect to $q_i$ and applying
Jacobi's formula Eq.~\eqref{eqn:jacobi}, together with the rank-one
structure of the adjugate matrix, yields
$
\mathcal{C}_\mathbf{q}^\mathrm{Cp}
\langle u^L_{1\mathbf{q}}|\partial_{q_i}
\mathbb{K}_{1\mathbf{q}}|u^R_{1\mathbf{q}}\rangle
+
\mathcal{C}_\mathbf{q}^\mathrm{Cp}
\langle u^L_{1\mathbf{q}}|
\partial_{z_{1\mathbf{q}}}
\mathbb{K}_{1\mathbf{q}}|u^R_{1\mathbf{q}}\rangle
\,\partial_{q_i}z_{1\mathbf{q}}
=0.
$
Evaluating this expression at $\mathbf{q}=\mathbf{0}$, where
$z_{1\mathbf{q}}=z_{1,-\mathbf{q}}$ implies
$\partial_{q_i}z_{1\mathbf{0}}=0$, gives the
first-order consistency condition
$
\langle u^L_{1\mathbf{0}}|
\partial_{q_i}\mathbb{K}_{1\mathbf{0}}
|u^R_{1\mathbf{0}}\rangle = 0.
$
Differentiating the secular equation once more with respect to $q_j$
and evaluating the result at $\mathbf{q}=\mathbf{0}$ yields
$
\operatorname{Tr}\!\big[
\frac{d\,\operatorname{adj}(\mathbb{K}_{1\mathbf{0}})}{dq_j}
\frac{d\mathbb{K}_{1\mathbf{0}}}{dq_i}
\big]
+
\operatorname{Tr}\!\big[
\operatorname{adj}(\mathbb{K}_{1\mathbf{0}})
\frac{d^2\mathbb{K}_{1\mathbf{0}}}{dq_i dq_j}
\big]
=0.
$
To evaluate the first trace term, we differentiate the rank-one
representation of the adjugate matrix and then evaluate the result at
$\mathbf{q}=\mathbf{0}$, obtaining
$
\frac{d\mathcal{C}_\mathbf{0}^\mathrm{Cp}}{dq_j}
|u^R_{1\mathbf{0}}\rangle
\langle u^L_{1\mathbf{0}}|
+
\mathcal{C}_\mathbf{0}^\mathrm{Cp}
|\partial_{q_j}u^R_{1\mathbf{0}}\rangle
\langle u^L_{1\mathbf{0}}|
+
\mathcal{C}_\mathbf{0}^\mathrm{Cp}
|u^R_{1\mathbf{0}}\rangle
\langle\partial_{q_j}u^L_{1\mathbf{0}}|.
$
Tracing this operator against
$
\frac{d\mathbb{K}_{1\mathbf{0}}}{dq_i}
=
\partial_{q_i}\mathbb{K}_{1\mathbf{0}},
$
which follows from $\partial_{q_i}z_{1\mathbf{0}}=0$, the term
proportional to
$
d\mathcal{C}_\mathbf{0}^\mathrm{Cp}/dq_j
$
vanishes identically by virtue of the first-order consistency condition.
Using the null-space derivative identities
$
\partial_{q_i}\mathbb{K}_{1\mathbf{0}}
|u^R_{1\mathbf{0}}\rangle
=
-\mathbb{K}_{1\mathbf{0}}
|\partial_{q_i}u^R_{1\mathbf{0}}\rangle
$
and
$
\langle u^L_{1\mathbf{0}}|
\partial_{q_i}\mathbb{K}_{1\mathbf{0}}
=
-\langle\partial_{q_i}u^L_{1\mathbf{0}}|
\mathbb{K}_{1\mathbf{0}},
$
the first trace contribution becomes
$
-\mathcal{C}_\mathbf{0}^\mathrm{Cp}
\langle\partial_{q_i}u^L_{1\mathbf{0}}|
\mathbb{K}_{1\mathbf{0}}
|\partial_{q_j}u^R_{1\mathbf{0}}\rangle
-
\mathcal{C}_\mathbf{0}^\mathrm{Cp}
\langle\partial_{q_j}u^L_{1\mathbf{0}}|
\mathbb{K}_{1\mathbf{0}}
|\partial_{q_i}u^R_{1\mathbf{0}}\rangle.
$
For the second trace term, all mixed contributions proportional to
$\partial_{q_i}z_{1\mathbf{0}}$ vanish at $\mathbf{q}=\mathbf{0}$, and the
total second derivative reduces to
$
\frac{d^2\mathbb{K}_{1\mathbf{0}}}{dq_i dq_j}
=
\partial^2_{q_iq_j}\mathbb{K}_{1\mathbf{0}}
+
\partial_{z_0}\mathbb{K}_{1\mathbf{0}}
(M^{-1}_{\mathrm{Cp}})_{ij}.
$
Substituting the rank-one representation of
$
\operatorname{adj}(\mathbb{K}_{1\mathbf{0}})
$
into the remaining trace then gives
$
\mathcal{C}_\mathbf{0}^\mathrm{Cp}
\langle u^L_{1\mathbf{0}}|
\partial^2_{q_iq_j}\mathbb{K}_{1\mathbf{0}}
|u^R_{1\mathbf{0}}\rangle
+
\mathcal{C}_\mathbf{0}^\mathrm{Cp}
\langle u^L_{1\mathbf{0}}|
\partial_{z_0}\mathbb{K}_{1\mathbf{0}}
|u^R_{1\mathbf{0}}\rangle
(M^{-1}_{\mathrm{Cp}})_{ij}.
$

To reveal the geometric structure of the Cooper problem, we invoke 
Eq.~\eqref{eq:K1q} at $\mathbf{q}=\mathbf{0}$, noting that the $\eta = 1$ 
contribution vanishes identically since $\chi_{1\mathbf{q}} = 0$ there. 
In direct analogy with the branch-resolved quantum metric of the two-body sector, 
we define the mode-resolved biorthogonal quantum-metric tensor 
as~\cite{montag26}
\begin{equation}
\begin{split}
g^{\eta'ij}_{\eta \mathbf{q}, \mathrm{Cp}} &=
\langle\partial_{q_i}u^L_{\eta\mathbf{q}}|u^R_{\eta'\mathbf{q}}\rangle
\langle u^L_{\eta'\mathbf{q}}|\partial_{q_j}u^R_{\eta\mathbf{q}}\rangle \\
&\quad +
\langle\partial_{q_j}u^L_{\eta\mathbf{q}}|u^R_{\eta'\mathbf{q}}\rangle
\langle u^L_{\eta'\mathbf{q}}|\partial_{q_i}u^R_{\eta\mathbf{q}}\rangle,
\label{eq: cooperbandresolved}
\end{split}
\end{equation}
for the $\eta$th collective mode, where $\eta \neq \eta'$. 
Inserting the spectral decomposition into the trace expression 
restricts the summation to the non-null sector $\eta \neq 1$, with each term 
weighted by the corresponding complex eigenvalue. Combining the resulting 
contributions and solving for the inverse mass tensor yields the central result 
of this section:
\begin{equation}
(M^{-1}_{\mathrm{Cp}})_{ij} =
\frac{
\langle u^L_{1\mathbf{0}}|\partial^2_{q_i q_j}\mathbb{K}_{1\mathbf{0}}|
u^R_{1\mathbf{0}}\rangle
-
\sum_{\eta\neq 1} \chi_{\eta\mathbf{0}}\,
g^{\eta ij}_{1 \mathbf{0}, \mathrm{Cp}}
}{
\langle u^L_{1\mathbf{0}}|
\partial_{-z_0} \mathbb{K}_{1\mathbf{0}}|
u^R_{1\mathbf{0}}\rangle
}.
\label{eq: Cooperinversemass}
\end{equation}
This expression is the direct analogue of Eq.~\eqref{eq: pairmass}.
Unlike Eq.~\eqref{eq: pairmass}, which is exact for the
two-body problem, Eq.~\eqref{eq: Cooperinversemass} is valid only within
the Gaussian approximation to the fluctuation action, and its inputs
$(T_c,\mu)$ require further approximations (Sec.~\ref{sec: sc}). 
An alternative, fully equivalent derivation of Eq.~\eqref{eq: Cooperinversemass}, 
based on on-shell differentiation of the secular equation~\eqref{eqn:secularK} 
and the pole condition~\eqref{eqn:poleC} rather than Jacobi's formula 
Eq.~\eqref{eqn:jacobi}, is provided in App.~\ref{sec:appB}.

Since $\mathbb{K}_{1\mathbf{q}}$ is non-Hermitian, both the numerator and
denominator of Eq.~\eqref{eq: Cooperinversemass} are generally complex
(at $T_c$, only the denominator).
Accordingly, the inverse mass tensor may be decomposed as
$
\mathbf{M}^{-1}_{\mathrm{Cp}} =
\operatorname{Re}\mathbf{M}^{-1}_{\mathrm{Cp}}
+ i\,\operatorname{Im}\mathbf{M}^{-1}_{\mathrm{Cp}}.
$
The quadratic dispersion in Eq.~\eqref{eq: mode} therefore acquires both
real and imaginary momentum-dependent contributions: the real part
governs the propagating character of the collective mode, whereas the
imaginary part determines its damping and 
lifetime~\cite{engelbrecht97, klimin21}, which arises from
the coupling of the collective mode to the continuum of fermionic pair
excitations.

Whenever $\omega_{1\mathbf{q}}\notin\mathcal{R}_{\mathbf{q}}$, as
in the BEC regime or, on a multiband lattice, when $2\mu$ lies in a gap
of the two-particle spectrum, $\mathcal{B}(\mathbf{q},\omega_{1\mathbf{q}})=0$
and $\mathbb{K}_{1\mathbf{q}}=\mathcal{A}(\mathbf{q},\omega_{1\mathbf{q}})$ is
Hermitian on shell, together with its on-shell derivatives, so that the
effective mass is purely real, and the collective mode is undamped. By
contrast, when $\mu$ lies inside a dispersive band, as in the
BCS regime, the continuum extends over the mode energy (over $\omega=0$
at $T_c$), allowing the collective mode to decay into fermionic pair
excitations. The resulting damping generates a finite anti-Hermitian
part of $\boldsymbol{\Gamma}^{-1}_{\downarrow}(\mathbf{q}, z)$ and, 
consequently, a finite imaginary part of $\mathbf{M}^{-1}_{\mathrm{Cp}}$. 
The crossover between these limits is continuous and controlled 
by the proximity of the collective-mode pole to the scattering 
threshold~\cite{sademelo93, klimin21}.

In this Hermitian regime, the biorthogonal formulation reduces smoothly
to the standard Hermitian one. In particular, the left and right 
eigenvectors coincide,
$
|u^L_{\eta\mathbf{q}}\rangle = |u^R_{\eta\mathbf{q}}\rangle
\equiv |u_{\eta\mathbf{q}}\rangle,
$
all eigenvalues $\chi_{\eta\mathbf{q}}$ are real, and the biorthogonal
mode-resolved quantum metric of Eq.~\eqref{eq: cooperbandresolved}
reduces to
$
g^{\eta' ij}_{\eta \mathbf{q}, \mathrm{Cp}} =
2\,\mathrm{Re}\,[
\langle\partial_{q_i}u_{\eta\mathbf{q}}|u_{\eta'\mathbf{q}}\rangle
\langle u_{\eta'\mathbf{q}}|\partial_{q_j}u_{\eta\mathbf{q}}\rangle
].
$
Under these conditions,
$\operatorname{Im}\mathbf{M}^{-1}_{\mathrm{Cp}}$
vanishes identically, and Eq.~\eqref{eq: Cooperinversemass} reduces to
the real inverse effective-mass tensor
\begin{equation}
(M^{-1}_{\mathrm{Cp}})_{ij} =
\frac{
\langle u_{1\mathbf{0}}|
\partial^2_{q_iq_j}\boldsymbol{\Gamma}^{-1}(\mathbf{0}, \omega_0)
|u_{1\mathbf{0}}\rangle
-
\sum_{\eta\neq 1}\chi_{\eta\mathbf{0}}\,g^{\eta ij}_{1\mathbf{0}, \mathrm{Cp}}
}{
\langle u_{1\mathbf{0}}|
\partial_{-\omega_0}\boldsymbol{\Gamma}^{-1}(\mathbf{0}, \omega_0)
|u_{1\mathbf{0}}\rangle
},
\label{eq: coopermass}
\end{equation}
where $\omega_{0} = z_{0}$ is real. This expression is formally identical
to the two-body result in Eq.~\eqref{eq: pairmass}: the first term
represents the ``conventional'' contribution arising from the momentum
curvature of the fluctuation propagator, while the second encodes the
quantum geometry of the many-body pairing manifold.
This structure also has a direct Ginzburg-Landau interpretation near
$T_c$~\cite{sademelo93, iskin23, kalkan2026}. By integrating out the massive
fluctuation modes of the Gaussian action, one obtains an effective theory
for the critical mode whose gradient coefficient is determined by the
numerator appearing in Eq.~\eqref{eq: coopermass}. In this formulation, the
geometric term arises from the momentum-dependent admixture of the
massive modes to the critical mode, providing a microscopic origin for
the pair quantum geometry appearing in the long-wavelength
Ginzburg-Landau theory. For instance, the associated coherence-length 
tensor is given by
$
(\xi_T^2)_{ij} = (\xi_{\mathrm{GL}}^2)_{ij} / \epsilon(T),
$
where $(\xi_{\mathrm{GL}}^2)_{ij}$ is given by the numerator of
Eq.~\eqref{eq: coopermass} divided by
$
2T_c
\langle u_{1\mathbf{0}}|
\partial_T\boldsymbol{\Gamma}^{-1}(\mathbf{0},0)
|u_{1\mathbf{0}}\rangle
$
evaluated at $T_c$, and $\epsilon(T)=(T_c-T)/T_c$.

At $T_c$, the critical mode has its root at
$(\mathbf{q},z)=(\mathbf{0},0)$, and the biorthogonal structure simplifies
even when this mode lies inside the continuum. Since
$\mathcal{B}(\mathbf{q},0)=0$, both sheets meet at $z=0$ in the Hermitian
static kernel $\boldsymbol{\Gamma}^{-1}_\downarrow(\mathbf{q},0)
=\boldsymbol{\Gamma}^{-1}(\mathbf{q},0)$, provided $\omega=0$ is not a
singular point of $\mathcal{B}$, and all momentum derivatives at fixed $z=0$
are therefore Hermitian. Since, in addition,
$\partial_{q_i}z_{1\mathbf{0}}=0$, the left and right eigenvectors and
their first momentum derivatives coincide at $\mathbf{q}=\mathbf{0}$, the
eigenvalues $\chi_{\eta\mathbf{0}}$ are real, and the metric of
Eq.~\eqref{eq: cooperbandresolved} reduces to the Hermitian one, so that
non-Hermiticity enters only through the frequency dependence. Above the
real axis, $\boldsymbol{\Gamma}^{-1}$ expands about $\mathcal{A}-i\mathcal{B}$, whereas
below it the first sheet expands about $\mathcal{A}+i\mathcal{B}$, and the
term $-2i\mathcal{B}(\mathbf{q},z)=-2iz\,\partial_\omega\mathcal{B}(\mathbf{q},0)
+\mathcal{O}(z^2)$ in Eq.~\eqref{eq:cont} restores the upper expansion,
so that near $(\mathbf{0},0)$
\begin{align}
\boldsymbol{\Gamma}^{-1}_\downarrow(\mathbf{q},z)
=\boldsymbol{\Gamma}^{-1}(\mathbf{q},0)
-z[\partial_{-\omega}\mathcal{A}(\mathbf{q},0)
+i\,\partial_\omega\mathcal{B}(\mathbf{q},0)],
\label{eq: GammaDownLinearZ}
\end{align}
up to $\mathcal{O}(z^2)$, 
which is all that the mass requires, while the dispersion at finite $\mathbf{q}$
follows from $\det\mathbb{K}_{1\mathbf{q}}=0$ with the full $z$ dependence
(Sec.~\ref{sec:cont}). With Eq.~\eqref{eq: GammaDownLinearZ},
Eq.~\eqref{eq: Cooperinversemass} becomes
\begin{equation}
(M^{-1}_{\mathrm{Cp}})_{ij} =
\frac{
\langle u_{1\mathbf{0}}|
\partial^2_{q_iq_j}\boldsymbol{\Gamma}^{-1}(\mathbf{0},0)
|u_{1\mathbf{0}}\rangle
-
\sum_{\eta\neq 1}\chi_{\eta\mathbf{0}}\,g^{\eta ij}_{1\mathbf{0}, \mathrm{Cp}}
}{
\langle u_{1\mathbf{0}}|
\partial_{-\omega}\mathcal{A}
+i\,\partial_{\omega}\mathcal{B}
|u_{1\mathbf{0}}\rangle
},
\label{eq: CooperMassTc}
\end{equation}
with $\mathcal{A}$ and $\mathcal{B}$ evaluated at
$(\mathbf{q},\omega)=(\mathbf{0},0)$. The numerator is the real
Ginzburg-Landau gradient coefficient of Eq.~\eqref{eq: coopermass},
positive semidefinite because the lowest eigenvalue of
$\boldsymbol{\Gamma}^{-1}(\mathbf{q},0)$ is minimal at $\mathbf{q}=\mathbf{0}$.
The dissipative term
$\langle u_{1\mathbf{0}}|\partial_\omega\mathcal{B}|u_{1\mathbf{0}}\rangle\geq0$
collects the two-fermion states with
$\xi_{n\mathbf{k}\uparrow}+\xi_{m,-\mathbf{k},\downarrow}=0$, i.e., at zero
energy relative to $2\mu$, into which a zero-energy pair can decay. It
vanishes when $0\notin\mathcal{R}_{\mathbf{0}}$, and
Eq.~\eqref{eq: CooperMassTc} then reduces to Eq.~\eqref{eq: coopermass}
with $\omega_0=0$. Since the numerator and the dissipative term are both
nonnegative, $\mathrm{Im}\,z_{1\mathbf{q}}\leq0$ to order $q^2$, and since
the numerator is common to the real and imaginary parts, dissipation
rescales the inverse mass by a complex scalar but leaves its geometric
content unchanged.

As in the two-body problem, the mode geometry contributes to
Eq.~\eqref{eq: Cooperinversemass} only if the sublattice texture of the
pairing fluctuations varies with $\mathbf{q}$. If
$|u_{1\mathbf{q}}\rangle = e^{i\gamma_\mathbf{q}}|u_1\rangle$ with a
$\mathbf{q}$-independent $|u_1\rangle$, as for uniform pairing
fluctuations $|u_1\rangle = (1,1,\ldots,1)^\mathrm{T}/\sqrt{N_b}$, then
$|\partial_{q_i}u_{1\mathbf{q}}\rangle$ is parallel to
$|u_{1\mathbf{q}}\rangle$ and the geometric terms
$g^{\eta ij}_{1\mathbf{0}, \mathrm{Cp}}$ vanish identically, because a
long-wavelength boost then modifies only the overall phase of the pairing
fluctuations. A momentum-dependent texture is therefore necessary for a
finite geometric contribution, but not sufficient, as discussed in
App.~\ref{sec:apppair}. The embedding dependence discussed below
Eq.~\eqref{eq: pairmass} applies to $\mathbb{K}_{1\mathbf{q}}$ as well.
In addition, when time-reversal symmetry is present and the pairing
fluctuations are uniform, the $\mathbf{q = 0}$ limit of
$\partial^2_{q_i q_j}\mathbb{K}_{1\mathbf{0}}$ reduces to the
``conventional'' contribution previously established in the 
literature~\cite{iskin23, iskin24}.

\subsection{Self-consistency relations for $T_c$ and $\mu$}
\label{sec: sc}

Here, we discuss the self-consistency equations for $T_c$ and $\mu$
in three dimensions, which serve as inputs for 
Eqs.~\eqref{eq: Cooperinversemass} and \eqref{eq: coopermass}
at a given filling $F$~\cite{nsr85, sademelo93, iskin24c}. 
The saddle-point condition follows from the generalized Thouless criterion,
$
\det \boldsymbol{\Gamma}^{-1}(\mathbf{q},0) = 0,
$
evaluated at $\mathbf{q} = \mathbf{0}$ for BCS-type pairing. In a multiband 
system, this condition determines the temperature at which the normal state 
becomes unstable toward superconductivity.
More specifically, the vanishing of the $\eta$th eigenvalue of the matrix
$
\boldsymbol{\Gamma}^{-1}(\mathbf{0},0)
$
at a temperature $T_{c_\eta}$ signals the onset of a gapless pairing mode. 
The physical transition temperature is then given by the highest of these 
values, i.e., 
$
T_c = \max\{T_{c_\eta}\},
$
and must be determined self-consistently together with the number equation.
For $T>T_c$, $\boldsymbol{\Gamma}^{-1}(\mathbf{0},0)\succ0$, and
since the instability is assumed to occur first at $\mathbf{q}=\mathbf{0}$,
$\boldsymbol{\Gamma}^{-1}(\mathbf{q},0)\succeq0$ for all $\mathbf{q}$ at
$T\geq T_c$, as used in Sec.~\ref{sec:Cp}.

The average particle filling per site,
$
F = \mathcal{N}/N
$
(with $0 \leq F \leq 2$), is obtained from the thermodynamic potential
$\Omega = \Omega_0 + \Omega_2$ via
$
F = -\partial_\mu \Omega/N.
$
Within this framework, the fluctuation contribution
$
\Omega_2 = T \sum_q \ln \det[T \boldsymbol{\Gamma}^{-1}(q)]
$
accounts for Gaussian-order fluctuations and provides a qualitatively
accurate description of the BCS-BEC crossover for all $U \neq 0$
\cite{sademelo93,nsr85}. Accordingly, the filling separates into a
saddle-point contribution $F_0$ and a fluctuation correction $F_2$.
The saddle-point term reads
$
F_0 = \frac{1}{N}\sum_{n\mathbf{k}\sigma}
f_{\mathrm{FD}}(\xi_{n\mathbf{k}\sigma}),
$
where
$
f_{\mathrm{FD}}(x) = \frac{1}{e^{x/T} + 1}
= \frac{1}{2}\left[
1 - \tanh\left(\frac{x}{2T}\right)
\right]
$
is the Fermi-Dirac distribution.
The fluctuation contribution further decomposes into a bound-state part
$F_{\mathrm{bs}}$, arising from the isolated poles of $\boldsymbol{\Gamma}(q)$,
and a scattering part $F_{\mathrm{sc}}$, associated with the branch cut of the
logarithm generated by the two-particle continuum. The bound-state contribution
is given by
$
F_{\mathrm{bs}} = \frac{2}{N} \sum_{\eta \mathbf{q}}
f_{\mathrm{BE}}(\omega_{\eta \mathbf{q}}),
$
where
$
f_{\mathrm{BE}}(x) = \frac{1}{e^{x/T} - 1}
= \frac{1}{2}\left[
\coth\left(\frac{x}{2T}\right) - 1
\right]
$
is the Bose-Einstein distribution, and $\omega_{\eta \mathbf{q}}$ denotes
the undamped collective-mode poles, i.e., the real roots of
Eq.~\eqref{eq:modesa}, measured from $2\mu$.

In the following, we restrict the analysis to the most relevant pole, 
which we denote by $\eta = 1$ without loss of generality, and
use its low-$\mathbf{q}$ expansion
\begin{equation}
\omega_{1 \mathbf{q}} 
\approx \frac{1}{2}\sum_{ij} (M^{-1}_\mathrm{Cp})_{ij} q_i q_j,\label{eq:wBq}
\end{equation}
where $\omega_{1\mathbf{0}}=0$ by the Thouless criterion.
We further neglect $F_{\mathrm{sc}}$ under the assumption that this pole 
remains spectrally well separated from the two-particle scattering continuum.
This is the case in the flat-band window of Sec.~\ref{sec:dmb},
where $2\mu$ lies in a gap of the two-particle spectrum and
$\omega_{1\mathbf{q}}$ is real. When the critical mode lies inside the
continuum, $z_{1\mathbf{q}}$ is complex, the pole no longer contributes to
$F_{\mathrm{bs}}$ in this form, and we instead determine $(T_c,\mu)$ from
the Thouless criterion and the mean-field number equation, $F = F_0$
(Sec.~\ref{sec:cont}).
We also note that $\omega_{1 \mathbf{q}}+2\mu$ may be
approximated by its two-body counterpart $E_{1 \mathbf{q}}$ in the
strong-coupling limit, where $2\mu$ approaches the two-body
threshold $E_{0}$. The formal justification for this replacement, namely,
that the many-body secular equation reduces exactly to the two-body
secular equation for $T \to 0$ with all bands empty, where
$\mathcal{X}_{n\mathbf{k}\sigma}\to1$, follows from the structural
correspondence established above between $\mathbb{K}_{1 \mathbf{q}}$ and 
$\mathbb{G}_{1 \mathbf{q}}$.
Under these approximations, the number equation reduces to
\begin{equation}
F \approx \frac{1}{N}\sum_{n\mathbf{k}\sigma}
f_{\mathrm{FD}}(\xi_{n\mathbf{k}\sigma})
+ \frac{2}{N}\sum_{\mathbf{q}}
f_{\mathrm{BE}}(\omega_{1 \mathbf{q}}),
\label{eq:numbereqn}
\end{equation}
where the factor of $2$ accounts for the two fermions comprising each
bosonic pair. The self-consistency cycle is then closed by solving
Eq.~\eqref{eq:numbereqn} together with the Thouless condition, thereby
determining $T_c$ and $\mu$ throughout the BCS-BEC crossover in 
three dimensions. Beyond the Gaussian action, this scheme
neglects $F_{\mathrm{sc}}$, keeps only the $\eta=1$ pole, and uses its
quadratic dispersion, Eq.~\eqref{eq:wBq}, and the many-body results of
Sec.~\ref{sec:dmb} are subject to these approximations.
There, we apply
it to a weakly interacting flat-band system, where the suppression of the
single-particle kinetic energy enhances the separation between the
collective pair mode and the fermionic scattering continuum.

\section{Numerical Illustrations}
\label{sec:ni}

\begin{figure*}[htb]
\centering
\includegraphics[width=\textwidth]{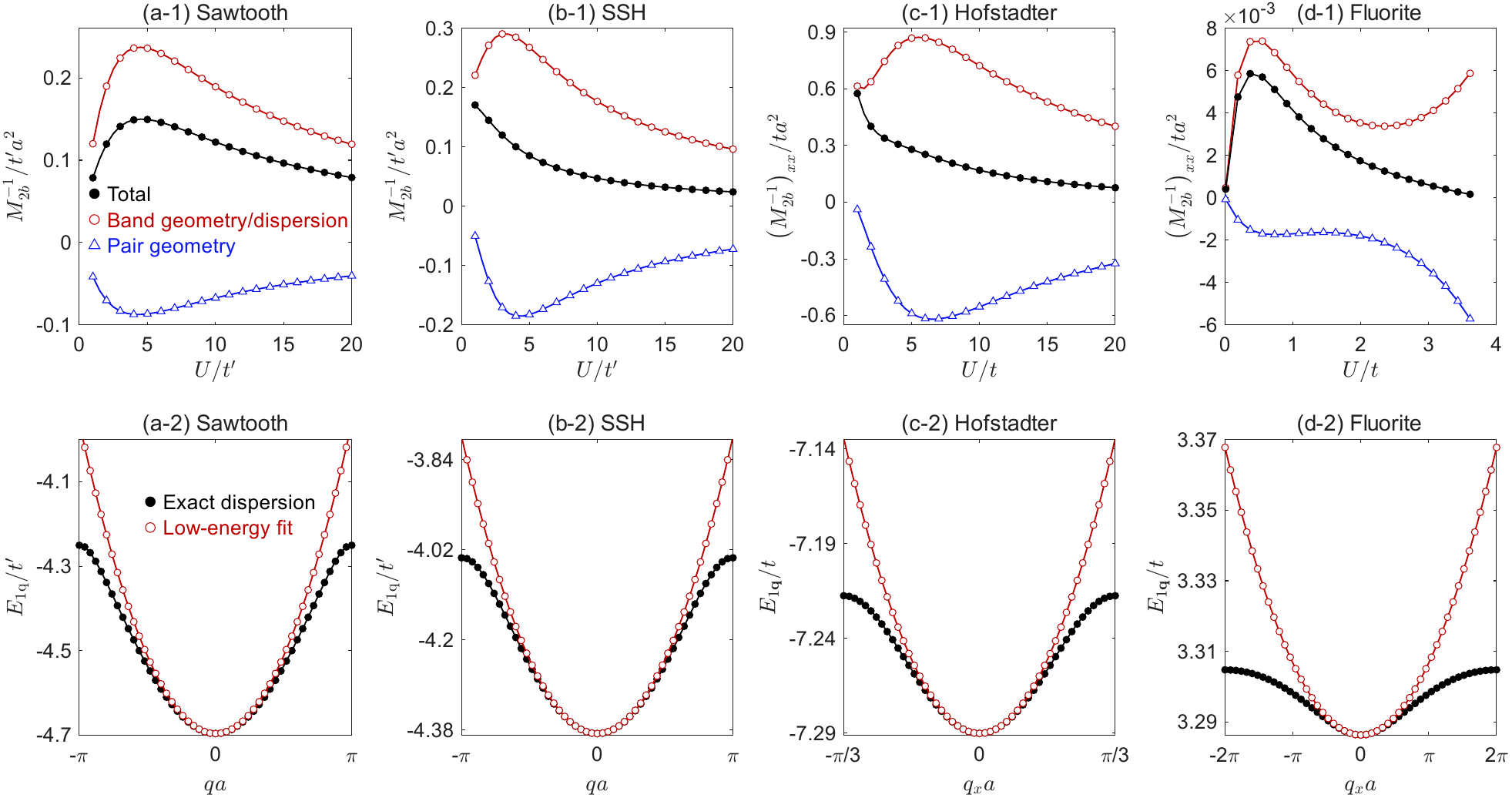}
\caption{
Decomposition of the pair inverse effective mass and validation
of the low-momentum dispersion for the (a) sawtooth, (b) SSH,
(c) Hofstadter, and (d) fluorite-like lattice models.
Top panels (a-1) through (d-1): ``conventional'' band-dispersion and
band-geometry contributions (red open circles), together with
the pair-geometry contribution (blue open triangles) to
Eq.~\eqref{eq: pairmass}, and their total sum
(black filled circles), shown as functions of the interaction strength.
Bottom panels (a-2) through (d-2): exact bound-state dispersions 
(black filled circles) compared with the quadratic expansion from
Eq.~\eqref{eq: pairmass} (red open circles), showing excellent agreement
in the small-$\mathbf{q}$ regime for representative parameters:
$U = 3t'$ (sawtooth and SSH models), $U = 5t$ (Hofstadter lattice), and
$U = t$ (fluorite-like lattice).
The lowest-lying bound state is shown for the sawtooth, SSH, and Hofstadter models.
For the fluorite-like model, the dispersion corresponds to the bound state
indicated by the black arrow in Fig.~\ref{fig:fullspectra}.
For the multidimensional Hofstadter and fluorite-like systems, the dispersion
is evaluated along the $q_x$ direction with all remaining momentum components
set to zero.
}
\label{fig:masspanel}
\end{figure*}

Having established the effective-mass theorems for the exact two-body problem
and the Gaussian fluctuations of the Cooper problem, we now investigate the
emergence of nontrivial pair quantum geometry in a diverse set of lattice
models: the sawtooth and Su-Schrieffer-Heeger (SSH) chains in one dimension,
the Hofstadter lattice in two dimensions, and the fluorite-like lattice in
three dimensions. The main numerical results are summarized in
Fig.~\ref{fig:masspanel}. As discussed in App.~\ref{sec:apppair}, the pairing
eigenvectors
$
|v_{1\mathbf{q}}\rangle
$
in these models have a sublattice texture that depends on
$\mathbf{q}$ already at low momenta, in the embeddings specified in
Sec.~\ref{sec:tbp}. In addition, the corresponding pair kernels
satisfy
$
\mathbb{G}_{1\mathbf{q}} \neq \mathbb{G}_{1,-\mathbf{q}},
$
revealing the absence of inversion symmetry in the pair sector.

\subsection{Sawtooth, SSH and Hofstadter lattices}
\label{sec:sl}

The sawtooth chain is a one-dimensional lattice with $N_b = 2$
sublattice degrees of freedom per unit cell, labeled by $S=\{A,B\}$,
and lattice constant $a$. The $A$ sublattice forms a uniform backbone
with nearest-neighbor hopping amplitudes
$
t_{A_j,A_i}=-t
$
for $j=i\pm1$. The $B$ sublattice occupies the apex positions of the
sawtooth structure, with each $B$ site coupled to its two neighboring
$A$ sites through
$
t_{B_i,A_i}=t_{B_j,A_i}=-t'
$
with $j=i-1$, while direct $B$-$B$ hopping is absent,
$
t_{B_j,B_i}=0.
$
After Fourier transformation, and in terms of the sublattice spinor
$
\boldsymbol{\psi}_{k\sigma}
=
(c_{Ak\sigma},c_{Bk\sigma})^\mathrm{T},
$
the Bloch Hamiltonian takes the pseudospin form
\begin{equation}
\mathbf{h}_{k\sigma}
=
d_k^0\tau_0+\mathbf{d}_k\cdot\boldsymbol{\tau}
\label{eq: sawtooth}
\end{equation}
where $\tau_0$ is the $2\times2$ identity matrix and
$
\boldsymbol{\tau}=(\tau_x,\tau_y,\tau_z)
$
denotes the Pauli matrices acting in the sublattice space.
The scalar and vector components are
$
d_k^0=t\cos(ka),
$
$
d_k^x=t'[1+\cos(ka)],
$
$
d_k^y=t'\sin(ka),
$
and
$
d_k^z=t\cos(ka).
$
The resulting band dispersions are
$
\varepsilon_{s k}=d_k^0+s d_k
$
with $s=\pm$, where
$
d_k=|\mathbf{d}_k|
=
\sqrt{
2 {t'}^2[1+\cos(ka)]
+t^2\cos^2(ka)
}.
$
The corresponding sublattice-projected Bloch amplitudes are
$
s_{A k\sigma}
=
\frac{-d_k^x+i d_k^y}
{\sqrt{2d_k(d_k-s d_k^z)}}
$
and
$
s_{B k\sigma}
=
\frac{d_k^z-s d_k}
{\sqrt{2d_k(d_k-s d_k^z)}
}.
$

The SSH model is the canonical two-band 
dimerized chain, characterized by
alternating intracell hopping $t$, which couples the $A$ and $B$ sites
within the same unit cell, and intercell hopping $t'$, which couples
the $B_i$ site to the $A_{i+1}$ site in the neighboring unit cell.
In the convention of Eq.~\eqref{eq: singleparticle}, the hopping
amplitudes are
$
t_{B_i,A_i}=-t
$
and
$
t_{A_{i+1},B_i}=-t',
$
so that the real-space Hamiltonian is
$
\mathcal{H}_{\sigma}
=
\sum_i \big(
t c^\dagger_{Bi\sigma}c_{A i\sigma}
+t' c^\dagger_{A,i+1,\sigma}c_{B i\sigma}
+\mathrm{H.c.}
\big),
$
where $\mathrm{H.c.}$ denotes the Hermitian conjugate.
The model possesses chiral (sublattice) symmetry,
$
\tau_z \mathbf{h}_{k\sigma} \tau_z
= -\mathbf{h}_{k\sigma},
$
which forbids a $\tau_z$ component in the Bloch Hamiltonian.
Consequently, the Hamiltonian lies entirely in the off-diagonal sector of
the $2\times2$ sublattice space:
\begin{equation}
\mathbf{h}_{k\sigma}
=
\begin{pmatrix}
0 & t+t' e^{-ika} \\
t+t' e^{ika} & 0
\end{pmatrix}
=
\mathbf{d}_k\cdot\boldsymbol{\tau},
\label{eq: ssh}
\end{equation}
where the nonzero components of the pseudospin field are
$
d_k^x=t+t'\cos(ka)
$
and
$
d_k^y=t'\sin(ka).
$
The band dispersions are
$
\varepsilon_{s k}
=
s\sqrt{
t^2+{t'}^2+2tt'\cos(ka)
}.
$
The corresponding sublattice-projected Bloch amplitudes are
$
s_{A k\sigma}
=
\frac{e^{i\varphi_k/2}}{\sqrt{2}}
$
and
$
s_{B k\sigma}
=
\frac{s\,e^{-i\varphi_k/2}}{\sqrt{2}},
$
where
$
\varphi_k=\arg(t+t' e^{-ika})
$
is the phase of the off-diagonal hopping element.

The Hofstadter model describes a two-dimensional square lattice with
lattice constant $a$ subjected to a uniform perpendicular magnetic field
$\mathbf{B}=B\hat{z}$. The field is incorporated through the Peierls
substitution,
$
t_{ij}\to t e^{i\theta_{ij}},
$
where
$
\theta_{ij}
=
\frac{2\pi}{\Phi_0}
\int_{\mathbf{r}_i}^{\mathbf{r}_j}
\mathbf{A}\cdot d\mathbf{l},
$
and $\Phi_0=h/e$ is the magnetic flux quantum. The relevant control
parameter is the magnetic flux per plaquette,
$
\Phi=Ba^2,
$
which is commonly expressed through the dimensionless ratio
$
\alpha=\Phi/\Phi_0.
$
For rational flux $\alpha=p/q$, with $p$ and $q$ coprime integers,
the magnetic field enlarges the unit cell by a factor of $q$ and
reduces the translational symmetry of the lattice.

Adopting the Landau gauge
$
\mathbf{A}=(0,Bx,0),
$
the hopping amplitudes along the $x$ direction remain real, while those
along the $y$ direction acquire position-dependent Peierls phases.
The tight-binding Hamiltonian is
$
\mathcal{H}_{\sigma}
=
-t\sum_{mn}
\big(
c^\dagger_{m+1,n \sigma}c_{mn\sigma}
+
e^{i2\pi\alpha m}
c^\dagger_{m,n+1,\sigma}c_{mn\sigma}
+
\mathrm{H.c.}
\big),
$
where $m$ and $n$ label lattice sites along the $x$ and $y$ directions,
respectively. For $\alpha=p/q$, the Hamiltonian is invariant under
translations by $q$ lattice spacings along the $x$ direction.
Consequently, the magnetic BZ is reduced to
$
k_x\in
\big[
-\frac{\pi}{qa},
\frac{\pi}{qa}
\big),
$
and
$
k_y\in
\big[
-\frac{\pi}{a},
\frac{\pi}{a}
\big).
$
After Fourier transformation in the magnetic unit-cell basis, the system
is described by the Harper equation, yielding the $q\times q$ Bloch
Hamiltonian
\begin{equation}
\mathbf{h}_{\mathbf{k}\sigma}
=
\begin{pmatrix}
a_{1,k_y} & -t & 0 & \cdots & 0 & -t e^{-iqak_x} \\
-t & a_{2,k_y} & -t & \cdots & 0 & 0 \\
0 & -t & a_{3,k_y} & \cdots & 0 & 0 \\
\vdots & \vdots & \vdots & \ddots & \vdots & \vdots \\
0 & 0 & 0 & \cdots & a_{q-1,k_y} & -t \\
-t e^{iqak_x} & 0 & 0 & \cdots & -t & a_{q,k_y}
\end{pmatrix},
\label{eq: hofstadter}
\end{equation}
where
$
a_{j,k_y}
=
-2t\cos[k_ya-2\pi\alpha(j-1)].
$
Diagonalization of $\mathbf{h}_{\mathbf{k}\sigma}$ yields
$q$ magnetic subbands. As a function of the flux ratio $\alpha$,
these subbands form the fractal energy spectrum known as the
Hofstadter butterfly.

\subsection{Fluorite-like lattice}
\label{sec:fl}

To extend our numerical analysis to a three-dimensional system with
nontrivial quantum geometry and an isolated flat band, we consider a
three-band tight-binding model defined on a fluorite-like 
lattice~\cite{neves24, duan24}.
The crystal basis contains $N_b=3$ sublattice sites per unit cell:
one $A$ site with on-site energy $\varepsilon_A$,
and two $B$ sites, labeled
$S=\{B_1,B_2\}$, each with on-site energy $\varepsilon_B$.
A key structural feature of the model is the complete absence of direct
hopping between the two $B$ sites, i.e.,
$
t_{B_1,B_2}=0.
$
Instead, inter-sublattice coupling occurs exclusively through two
complex, direction-dependent $A$-$B$ hopping channels characterized by
the amplitudes $t_1$ and $t_2$.
After Fourier transformation to momentum space, and in terms of the
three-component sublattice spinor
$
\boldsymbol{\psi}_{\mathbf{k}\sigma}
= (c_{A\mathbf{k}\sigma}, c_{B_1\mathbf{k}\sigma},
c_{B_2\mathbf{k}\sigma})^\mathrm{T},
$
the $3\times3$ Bloch Hamiltonian takes the form
\begin{equation}
\mathbf{h}_{\mathbf{k}\sigma} =
\begin{pmatrix}
\varepsilon_A & f_{1,\mathbf{k}} & f_{2,\mathbf{k}} \\
f_{1,\mathbf{k}}^* & \varepsilon_B & 0 \\
f_{2,\mathbf{k}}^* & 0 & \varepsilon_B
\end{pmatrix},
\label{eq: fluorite}
\end{equation}
where the off-diagonal matrix elements are
$
f_{1,\mathbf{k}} = -t_1 e^{ia(k_x+k_y+k_z)/4},
$
and
$
f_{2,\mathbf{k}} = -t_2\big(e^{ia(k_x+k_y+k_z)/4}+e^{ia(k_x-k_y-k_z)/4}
+ e^{ia(-k_x+k_y-k_z)/4} + e^{ia(-k_x-k_y+k_z)/4} \big).
$

\begin{figure}[htb]
\centering
\includegraphics[width=0.65\columnwidth]{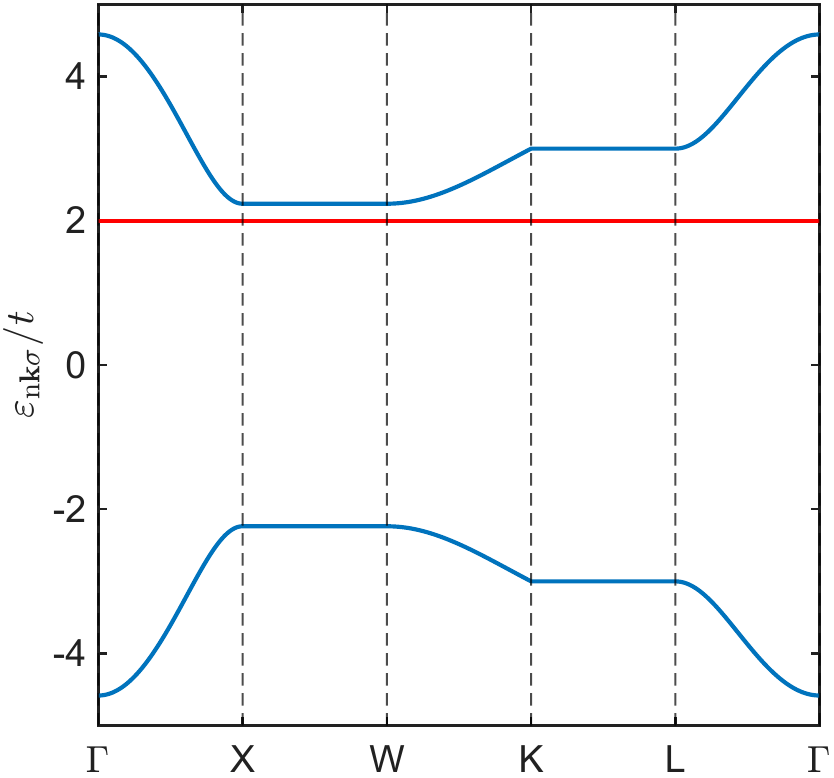}
\caption{Energy spectrum of the fluorite-like lattice for parameters
$t_1 = t_2 = t$ and $\varepsilon_A = -\varepsilon_B =  -2t$.
The Bloch bands are plotted along a high-symmetry path in the
first BZ, with coordinates defined as $\Gamma = (0, 0, 0)$,
$X = (0, 0, 2\pi/a)$, $W = (\pi/a, 0, 2\pi/a)$,
$K = (\pi/2a, \pi/2a, 2\pi/a)$, and $L = (\pi/a, \pi/a, \pi/a)$.}
\label{fig:spectrafluorite}
\end{figure}

This model provides an ideal platform for our numerical framework,
as its bipartite-like connectivity generates destructive quantum
interference that isolates a localized state. As a result, a
macroscopically degenerate flat band with energy $\varepsilon_B$
emerges throughout the Brillouin zone. The remaining spectrum consists
of two dispersive bands symmetric about the average on-site potential,
with eigenvalues
$
\varepsilon_{\pm,\mathbf{k}} = \frac{\varepsilon_A+\varepsilon_B}{2}
\pm \sqrt{\left(\frac{\varepsilon_A-\varepsilon_B}{2}\right)^2
+ |f_{1,\mathbf{k}}|^2 + |f_{2,\mathbf{k}}|^2 }.
$
Here, the dispersive contribution depends only on the combined hopping
magnitude
$
|f_{1,\mathbf{k}}|^2+|f_{2,\mathbf{k}}|^2,
$
which evaluates to
$
t_1^2 + 4t_2^2 \big[ 1 + \cos(k_xa/2)\cos(k_ya/2) + \cos(k_xa/2)\cos(k_za/2)
+ \cos(k_ya/2)\cos(k_za/2) \big].
$
The resulting band structure, consisting of an isolated flat band
bounded by upper and lower dispersive branches, is shown along a
high-symmetry path of the first Brillouin zone in
Fig.~\ref{fig:spectrafluorite} for the representative parameter set
$t_1 = t_2 = t$ and $\varepsilon_A = -\varepsilon_B = -2t$.

\begin{figure}[htb]
\centering
\includegraphics[width=\columnwidth]{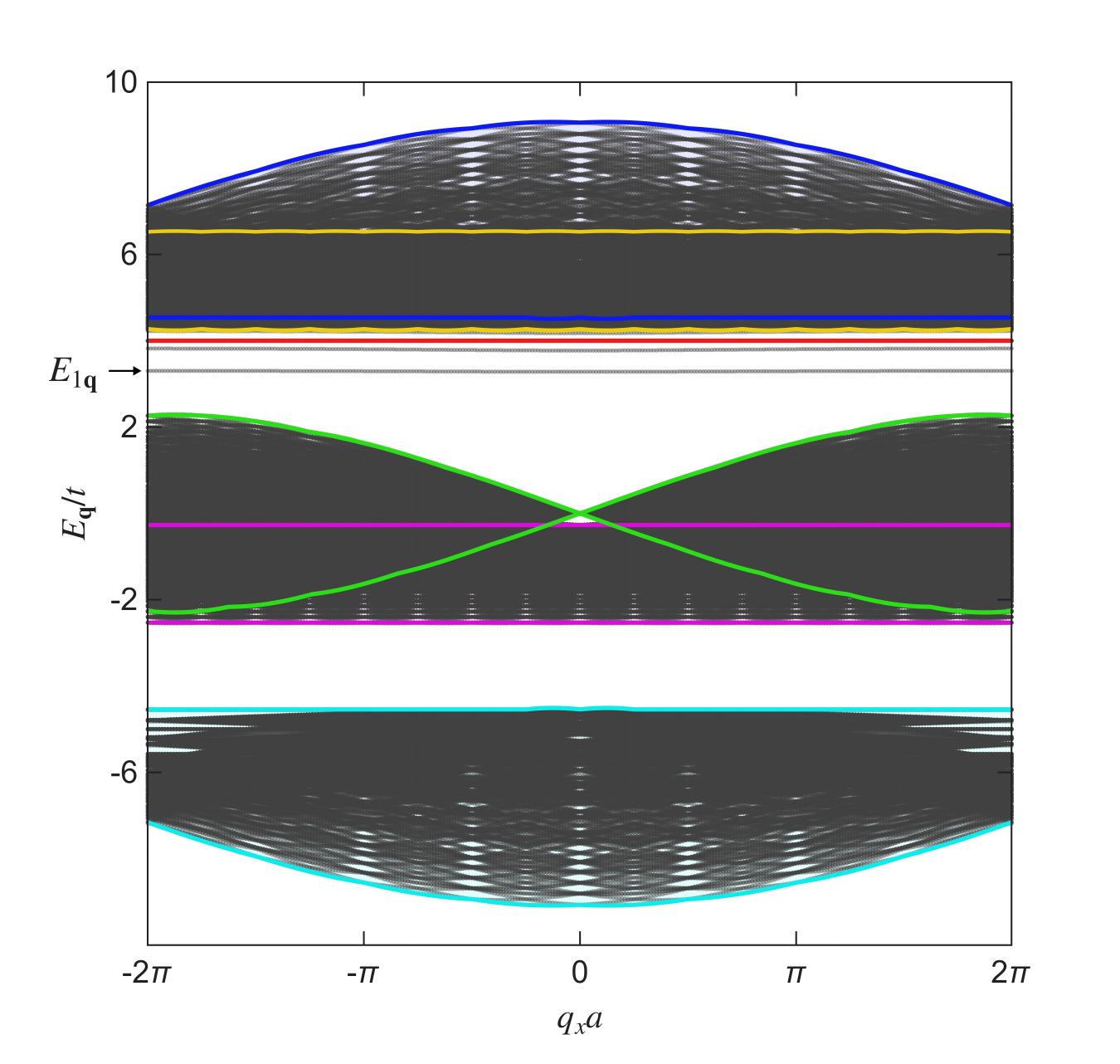}
\caption{
Two-body energy spectrum of the fluorite-like lattice for
$t_1 = t_2 = t$ and $\varepsilon_A = -\varepsilon_B =-2t$,
obtained by exact diagonalization of Eq.~\eqref{eq: twobody} at
interaction strength $U = t$ using $N_c = 100$ momentum points.
The horizontal red line denotes the dispersionless flat-band continuum
at energy $2\varepsilon_B$. Immediately below it, two discrete bound
states emerge within the spectral gap. The lower of these, indicated by
the black arrow, is the state analyzed in Fig.~\ref{fig:masspanel}(d)
through its effective mass. This is also the relevant bound state for
the Cooper problem when $\mu$ lies in the flat Bloch band of the 
noninteracting system. The remaining colored curves trace
the upper and lower edges
$
\max(\varepsilon_{n\mathbf{k}\uparrow}
+ \varepsilon_{m,-\mathbf{k}+\mathbf{q}, \downarrow})
$
and
$
\min(\varepsilon_{n\mathbf{k}\uparrow}
+ \varepsilon_{m,-\mathbf{k}+\mathbf{q}, \downarrow}),
$
respectively, of each scattering channel $(n,m)$, serving as guides to
the eye for the continuum structure.
}
\label{fig:fullspectra}
\end{figure}

The two-body energy spectrum of the fluorite-like lattice, shown in
Fig.~\ref{fig:fullspectra}, is obtained by exact diagonalization of
Eq.~\eqref{eq: twobody} at each center-of-mass momentum $q_x$ for
$t_1 = t_2 = t$ and $\varepsilon_A = -\varepsilon_B = -2t$
at interaction strength $U = t$. For a system with $N_b = 3$
single-particle bands and $N_c = 100$ momentum points, the resulting
eigenvalue problem has dimension $900$ at each $q_x$, providing a dense
sampling of the continuum structure.
The spectrum reflects the multiband nature of the underlying
single-particle Hamiltonian. The scattering continua, constructed from
all pairwise combinations of single-particle eigenstates,
$
\varepsilon_{n\mathbf{k}\uparrow} 
+ \varepsilon_{m,-\mathbf{k}+\mathbf{q}, \downarrow},
$
span a broad energy window whose upper and lower boundaries are traced
by colored guide lines corresponding to distinct channels $(n,m)$. The
dispersionless red line identifies the flat-band continuum, fixed at
energy $2\varepsilon_B$, which follows directly from
$
\varepsilon_{\mathrm{flat},\mathbf{k}\uparrow}
+ \varepsilon_{\mathrm{flat},-\mathbf{k}+\mathbf{q},\downarrow}
= 2\varepsilon_B.
$
Below the flat-band continuum, two discrete bound states appear within
the spectral gap separating it from the lower dispersive continuum. 
Among them, we focus on the lowest-lying state, indicated by
the black arrow in Fig.~\ref{fig:fullspectra}, and track its evolution
to extract the pair effective mass as a function of interaction strength
up to $U = 3.6t$. In this regime, the state remains well isolated within
the gap. Upon further increasing $U$, it moves toward the lower
scattering continuum and eventually loses its spectral isolation.

\subsection{Two-body discussion}
\label{sec:d}

Figure~\ref{fig:masspanel} reveals a unified picture of how pair quantum
geometry competes with ``conventional'' band-dispersion and band-geometry
effects in determining the inverse effective mass across one-, two-, and
three-dimensional multiorbital lattices.
A useful starting point is the strong-coupling limit. Since the matrix
elements of $\mathbb{G}_{1\mathbf{0}}$ contain the denominators
$
\varepsilon_{n\mathbf{k}\uparrow}
+\varepsilon_{m,-\mathbf{k+q},\downarrow}
-E_0,
$
and $E_0\to -U$ for large attraction, one finds
$\mathbb{G}_{1\mathbf{0}}\sim 1/U$.
Using
$
\mathbb{G}_{1\mathbf{0}}
|\partial_{q_i}v_{1\mathbf{0}}\rangle
=
-\partial_{q_i}\mathbb{G}_{1\mathbf{0}}
|v_{1\mathbf{0}}\rangle,
$
together with
$\partial_{q_i}\mathbb{G}_{1\mathbf{0}}\sim t/U^2$,
it follows that
$|\partial_{q_i}v_{1\mathbf{0}}\rangle\sim a t/U$.
As a result, both the pair-geometric and ``conventional'' terms in
Eq.~\eqref{eq: pairmass} scale as $a^2 t^2/U^3$, while
$\partial_{-E_0}\mathbb{G}_{1\mathbf{0}}\sim 1/U^2$.
Consequently, both contributions to the inverse mass tensor decay
universally as
$
(M^{-1}_{2b})_{ij}\sim a^2 t^2/U.
$
This scaling has a simple physical interpretation. As the attraction 
strength increases, the bound pair becomes increasingly localized 
in real space, reducing its sensitivity to the momentum-space 
structure of the underlying lattice. The inverse effective mass 
therefore decreases and ultimately vanishes in the atomic limit. 
While this overall strong-coupling scaling is universal, the 
relative importance of the ``conventional'' and pair-geometric 
contributions remains strongly model dependent. This behavior is 
clearly seen in the lowest-lying two-body bound-state branch of 
the sawtooth, SSH, and Hofstadter lattices, where both contributions 
eventually exhibit the same strong-coupling suppression. 
The fluorite-like lattice provides a striking exception: 
the proximity of a dispersive continuum keeps the pair sensitive 
to the underlying multiband structure and prevents the geometric 
contribution from being quenched even at comparatively strong coupling.

\textit{Sawtooth and SSH lattices} are shown in Figs.~\ref{fig:masspanel}(a-1),
(a-2), and (b-1), (b-2). For $t'=\sqrt{3}t$, the lowest-lying bound
states of the two lattices exhibit a broadly similar evolution, illustrating
the competition between the ``conventional'' and pair-geometric
contributions. The ``conventional'' contribution initially increases with
attraction, reaches a maximum near $U\approx3t'$, and then decreases at
strong coupling, while the pair-geometry contribution remains negative
and reaches its largest magnitude around $U\approx4t'$-$4.5t'$. The total
inverse mass therefore remains positive, with a monotonic decrease in the
SSH lattice and a non-monotonic behavior in the sawtooth lattice. The SSH
lattice exhibits a somewhat larger geometric correction. Notably,
chiral symmetry enforces $|\beta_{A\mathbf{q}}|=|\beta_{B\mathbf{q}}|$ for
every $\mathbf{q}$ in this model, so that its pair geometry originates
entirely from the momentum-dependent relative phase of
$\beta_{B\mathbf{q}}/\beta_{A\mathbf{q}}$, in line with the condition
discussed in App.~\ref{sec:apppair}, namely, that a uniform magnitude of the pairing
amplitudes does not imply a vanishing pair geometry. The excellent
agreement between the exact dispersions and their quadratic expansions at
$U=3t'$ confirms the accuracy of Eq.~\eqref{eq: boundstate}.

\textit{Hofstadter lattice} is shown in Figs.~\ref{fig:masspanel}(c-1)
and (c-2). For $\alpha=1/3$, the lowest-lying bound state exhibits the same
overall competition despite the nontrivial magnetic-band structure. The
pair-geometry contribution reaches a maximum magnitude of approximately
$0.62ta^2$ around $U\approx7t$ before decreasing at stronger coupling.
Together with the one-dimensional results, this demonstrates that pair
quantum geometry is not tied to a particular dimensionality or band
structure, but arises from the internal geometry of the paired-state
manifold. The quadratic expansion at $U=5t$ again accurately reproduces
the low-momentum dispersion.

\textit{Fluorite-like lattice} is shown in Figs.~\ref{fig:masspanel}(d-1)
and (d-2). Its behavior is qualitatively different because the relevant
bound state lies in the spectral gap between the flat-band continuum and
a nearby dispersive continuum (see Fig.~\ref{fig:fullspectra}). Increasing
$U$ drives the bound state toward the adjacent continuum, strongly
modifying the energy denominators entering $\mathbb{G}_{1\mathbf{0}}$.
As a result, both the ``conventional'' and total contributions initially rise
sharply, while the pair-geometric contribution develops a non-monotonic
interaction dependence and remains quantitatively important into the
strong-coupling regime. Thus, nearby continuum states can prevent the
usual suppression of pair geometry observed in the lowest-lying bound 
states of sawtooth, SSH, and Hofstadter lattices. The excellent agreement 
between the exact dispersion and its quadratic expansion at $U=t$ confirms 
the validity of Eq.~\eqref{eq: boundstate} in this three-dimensional 
multiorbital setting.

\begin{figure*}[htb]
\centering
\includegraphics[width=\textwidth]{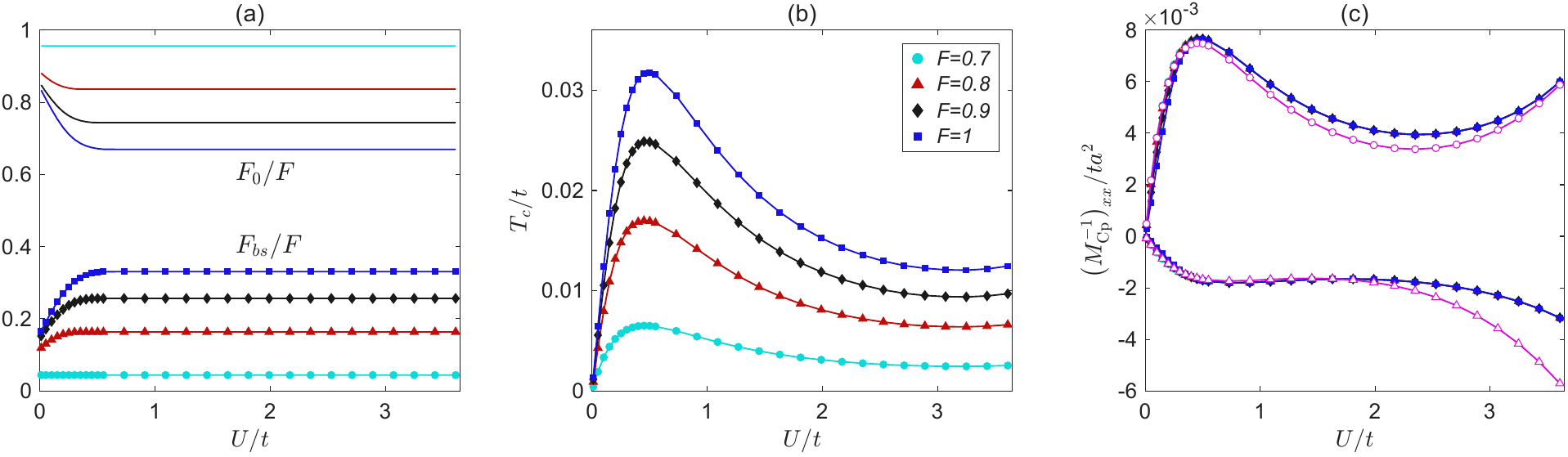}
\caption{
(a-b) Self-consistent solution of the number equation~\eqref{eq:numbereqn}
together with the Thouless criterion for $T_c$ on the fluorite-like lattice
for filling factors $F=0.7,0.8,0.9,1$.
(a) Bound-state fraction $F_{\mathrm{bs}}/F$ (line plots with data points)
and saddle-point fraction $F_0/F$ (line plots) as functions of interaction
strength $U/t$.
(b) Corresponding critical temperature $T_c/t$ as a function of $U/t$.
(c) $xx$-component of the Cooper-pair inverse effective mass,
$(M_{\mathrm{Cp}}^{-1})_{xx}$, obtained from Eq.~\eqref{eq: coopermass}.
The ``conventional'' contribution and pair-geometry contribution are nearly
independent of $F$, and they are compared with the two-body result,
$(M_{2b}^{-1})_{xx}$ (magenta data points, shown in red and blue in Fig.~\ref{fig:masspanel}(d-1)).
}
\label{fig:masspanelsc}
\end{figure*}
\begin{figure*}[htb]
\centering
\includegraphics[width=\textwidth]{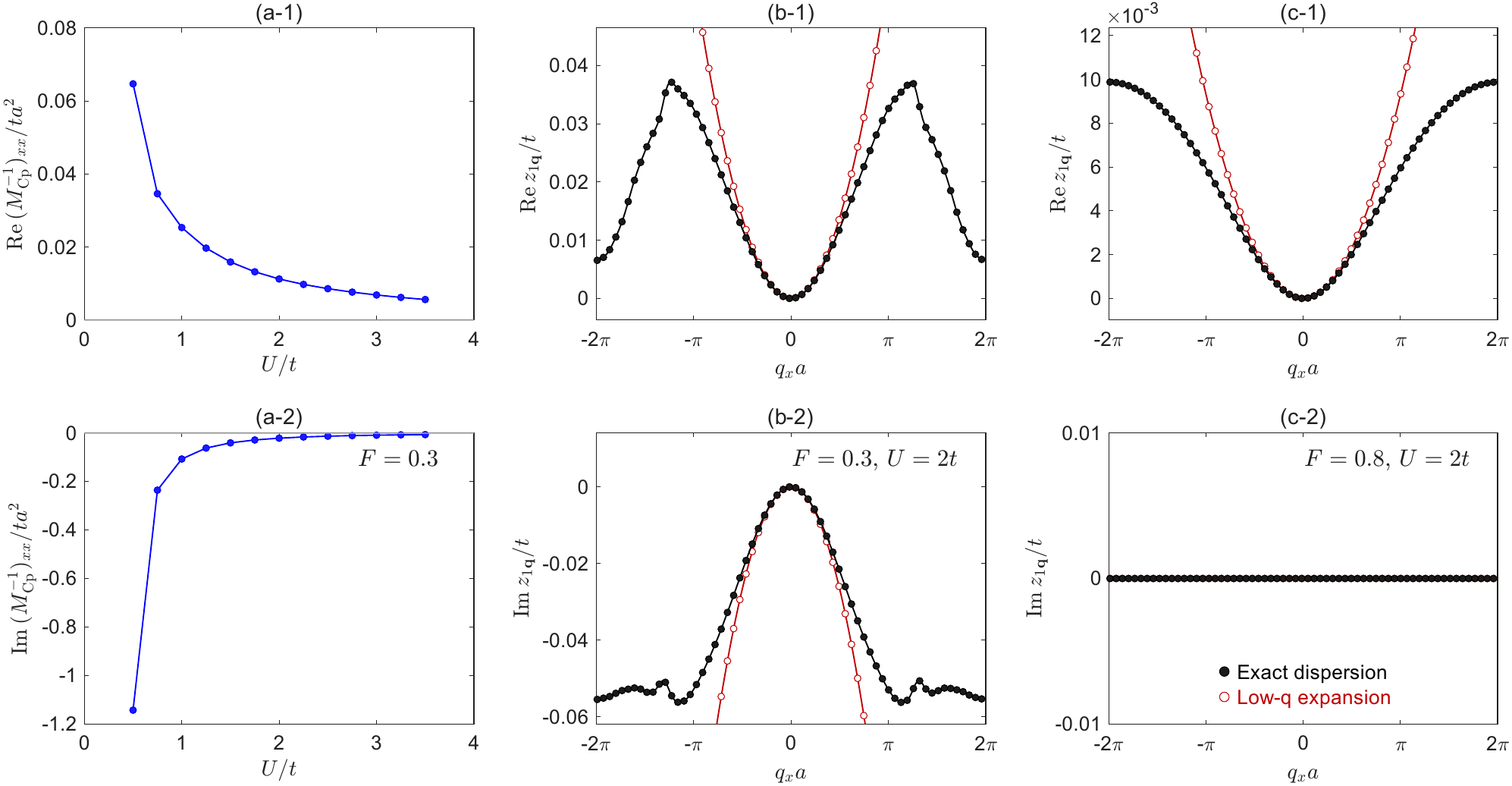}
\caption{
Cooper pair at $T_{c}$ for the fluorite-like lattice, with $(T_{c},\mu)$ from the
Thouless criterion and the mean-field number equation.
(a-1) Real and (a-2) imaginary parts of $(M^{-1}_{\mathrm{Cp}})_{xx}$,
Eq.~\eqref{eq: CooperMassTc}, as functions of $U/t$ for $F=0.3$ ($\mu$ in the lower
dispersive band).
(b-1) Real and (b-2) imaginary parts of $z_{1\mathbf{q}}$ along $q_x$ for $F=0.3$
and $U=2t$, where the mode lies inside the two-particle continuum.
(c-1) and (c-2) Same as (b-1) and (b-2) for $F=0.8$ ($\mu$ just below the
partially occupied flat band), where the mode lies in a gap of the continuum and
$z_{1\mathbf{q}}$ is real.
Black filled circles: roots of
$\det\boldsymbol{\Gamma}^{-1}_\downarrow(\mathbf{q},z)=0$ with the full $z$
dependence; red open circles: low-$q$ expansion
$\frac{1}{2}(M^{-1}_{\mathrm{Cp}})_{xx}q_x^2$.
}
\label{fig:complexmass}
\end{figure*}
\subsection{Many-body discussion}
\label{sec:dmb}

To connect our two-body results to the physically more relevant
many-body problem, we solve the self-consistency equations
of Sec.~\ref{sec: sc} for the fluorite-like lattice at filling factors
$F = \{0.7, 0.8, 0.9, 1\}$. Given the three-band structure shown in
Fig.~\ref{fig:spectrafluorite}, complete filling of the lower dispersive
band corresponds to $F = 2/3$ (i.e., $1/3$ per spin). The filling factors
considered here therefore lie just above this value, placing $\mu$
within the flat-band manifold $2/3 < F \le 4/3$, where $F=1$ corresponds
to a half-filled flat band.

Figure~\ref{fig:masspanelsc}(a) shows the decomposition of the filling into
unpaired-fermion ($F_0/F$) and bound-pair ($F_\mathrm{bs}/F$) fractions as
a function of interaction strength $U/t$. Owing to the flat-band manifold,
bound-pair formation becomes substantial already at weak coupling
($U \sim 0.5t$), where virtually all particles above the fully filled
lower dispersive band are converted into pairs. Consequently,
the lower dispersive band remains fully occupied by unpaired fermions,
$F_0 \rightarrow 2/3$, while the excess filling in the flat-band
manifold is carried almost entirely by bound pairs,
$F_\mathrm{bs} \rightarrow F-2/3$. The corresponding critical temperature
$T_c$ is shown in Fig.~\ref{fig:masspanelsc}(b) as a function of $U/t$.
In the weak-coupling regime, $T_c$ increases approximately linearly
with $U/t$, characteristic of flat-band pairing, before reaching a
maximum at intermediate coupling and subsequently
decreasing in the strong-coupling regime. Moreover, increasing the
flat-band filling fraction systematically enhances $T_c$
across the entire interaction range.

Using the self-consistent $(T_c, \mu)$ values, we evaluate the Cooper-pair 
inverse effective mass from Eq.~\eqref{eq: coopermass}, which is 
shown in Fig.~\ref{fig:masspanelsc}(c).
As in Fig.~\ref{fig:masspanel}, the inverse effective mass is decomposed
into ``conventional'' and pair-geometric contributions, with the 
two-body result from Fig.~\ref{fig:masspanel}(d-1) shown for comparison.
The two contributions exhibit nearly identical behavior across all four
filling factors, indicating that the qualitative structure of the
decomposition is largely insensitive to filling within this range.
The ``conventional'' contribution closely follows its two-body counterpart
over most of the interaction range, with some deviation emerging at
intermediate coupling, while retaining the same overall trend.
The pair-geometric contribution likewise follows the qualitative
behavior of its two-body counterpart, with a small deviation emerging
at strong coupling. This deviation is consistent with the fact that
$\omega_{1\mathbf q}+2\mu\approx E_{1\mathbf q}$ becomes exact
only when $\mathcal{X}_{n\mathbf{k}\sigma}\to1$ for all states, i.e., when
all bands are empty, in which case $\omega_{1\mathbf 0}=0$ gives
$2\mu=E_0$, whereas in the self-consistent many-body calculation the
lower dispersive band is filled and $\mu$ lies within a few $T_c$ of the
flat band, so that $\mathcal{X}_{n\mathbf{k}\sigma}\neq1$.

\subsection{Cooper pairs inside the continuum}
\label{sec:cont}

In the flat-band window of Sec.~\ref{sec:dmb}, $0\notin\mathcal{R}_{\mathbf{0}}$
and Eq.~\eqref{eq: CooperMassTc} reduces to the real mass of
Eq.~\eqref{eq: coopermass}. To illustrate the complex mass, we place $\mu$ in
the lower dispersive band ($F=0.3$), where $0\in\mathcal{R}_{\mathbf{0}}$, and
contrast it with $F=0.8$, where $\mu$ lies between the lower and the flat band.
Since the bound-pair approximation of Sec.~\ref{sec: sc} requires an undamped
pole, we determine $(T_c,\mu)$ for both fillings from the Thouless criterion
and the mean-field number equation, $F=F_0$. The continuation of
Eq.~\eqref{eq:cont} is evaluated by broadening the $\delta$ functions in
Eq.~\eqref{eq:AB} by a Gaussian of width $\delta\omega=0.08t$, which turns each pole
$1/(z-\xi_{n\mathbf{k}\uparrow}-\xi_{m,-\mathbf{k}+\mathbf{q},\downarrow})$ into a
Faddeeva function that is analytic in the whole complex plane, on a grid of
$30^3$ momenta in the fcc Brillouin zone.

For $F=0.3$ [Figs.~\ref{fig:complexmass}(a-1) and (a-2)],
$\mathrm{Re}\,(M^{-1}_{\mathrm{Cp}})_{xx}>0$ and
$\mathrm{Im}\,(M^{-1}_{\mathrm{Cp}})_{xx}<0$ for all $U$, consistent with
$\mathrm{Im}\,z_{1\mathbf{q}}\leq0$. The imaginary part dominates at weak
coupling, where the Cooper pair relaxes diffusively rather than propagates,
and the two parts become comparable as $U$ increases. At $U=2t$
[Figs.~\ref{fig:complexmass}(b-1) and (b-2)], $z_{1\mathbf{q}}$ is complex up
to the zone boundary $X$ ($q_xa=\pm2\pi$). The low-$q$ expansion
$z_{1\mathbf{q}}\approx\frac{1}{2}(M^{-1}_{\mathrm{Cp}})_{xx}q_x^2$, which
follows from Eq.~\eqref{eq: CooperMassTc} with $z_{1\mathbf{0}}=0$, reproduces
the exact root at long wavelengths. At larger $q$,
$|\mathrm{Im}\,z_{1\mathbf{q}}|$ saturates at a value comparable to $\delta\omega$, 
so that the large-$q$ structure is only qualitative.
For $F=0.8$ [Figs.~\ref{fig:complexmass}(c-1) and (c-2)], $\omega=0$ lies, for
all $\mathbf{q}$, in a gap between the continuum of lower-band/upper-band pairs and that
of flat-band/upper-band pairs, while the flat-band pairs form an isolated pole at $2(2t-\mu)$
rather than a branch cut. Hence $\mathcal{B}=0$ near $\omega=0$, and both the
mass and $z_{1\mathbf{q}}$ are real. Thus, on a multiband lattice, an undamped
Cooper-pair mode with a real mass arises whenever $2\mu$ lies in a gap of the
two-particle spectrum, without reaching the BEC regime.

\section{Conclusion}
\label{sec:conc}

In this work, we uncovered a geometric hierarchy governing the effective
masses of single particles, two-body bound states, and Cooper pairs in
multiband systems, all three following from the isolated zero
eigenvalue of a kernel. While the effective mass of an individual
particle is known to be influenced by the quantum geometry of Bloch
states, we showed that paired states possess an additional geometric
structure associated with the pairing manifold. This \emph{pair quantum
geometry} is distinct from the underlying band geometry and enters the
effective mass through a quantum metric constructed from the eigenvectors
of the pairing kernel. As for the band quantum metric, the
separation into ``conventional'' and pair-geometric terms depends on the
orbital embedding, whereas the total mass does not.

For the two-body problem in vacuum, we derived an exact effective-mass
theorem that separates the inverse effective-mass tensor into
``conventional'' and pair-geometric contributions. Exact numerical
calculations for the sawtooth, SSH, Hofstadter, and fluorite-like lattices,
spanning one, two, and three dimensions, demonstrated that the geometric
contribution can be quantitatively significant over a broad range of
interaction strengths. We further established the corresponding
many-body theorem for Cooper pairs near $T_c$ within Gaussian fluctuation
theory. Self-consistent calculations in the flat-band regime of
the fluorite-like lattice show that the ``conventional'' and
pair-geometric contributions follow their two-body counterparts, with
small deviations at stronger coupling, so that the geometric effects
persist in the many-body problem near $T_c$. When the collective mode lies
inside the two-particle continuum, the analytic continuation of the
fluctuation kernel to the second Riemann sheet makes it non-Hermitian,
which requires a biorthogonal extension of the geometric formalism and
yields a complex Cooper-pair effective mass, whose real and imaginary
parts characterize propagation and damping, respectively. At $T_c$, the
inverse mass is evaluated at $(\mathbf{q},z)=(\mathbf{0},0)$, where the
continued kernel reduces to the Hermitian static kernel. Its geometric
numerator is therefore real, and dissipation enters only through the
complex frequency derivative in the denominator, which rescales the
inverse mass by a complex scalar without changing its pair-geometric
content. The full biorthogonal structure is required at finite
$\mathbf{q}$, where $z_{1\mathbf{q}}$ is complex. For the fluorite-like
lattice, the complex mass is predominantly imaginary at weak coupling and
reproduces the exact damped dispersion at long wavelengths, whereas an
undamped mode with a real mass arises when $2\mu$ lies in a gap of the
two-particle spectrum, without reaching the BEC regime.

An important consequence is that pair quantum geometry can enter a broader
class of superconducting observables governed at long wavelengths by the
Cooper-pair effective mass. In particular, within Ginzburg-Landau theory,
the coherence length, penetration depth, superfluid weight, and upper
critical field are related to the gradient coefficient of the critical
mode. By integrating out the massive modes of the Gaussian action, we have
verified that the resulting effective Ginzburg-Landau theory provides an
alternative derivation of the Cooper-pair effective-mass
theorem, in line with the construction of Ref.~\cite{kalkan2026}. The
pair-geometric contribution to the gradient coefficient originates from
the virtual admixture of the massive modes to the critical mode, providing
a microscopic origin for geometric corrections to these long-wavelength
observables.

Taken together, our results show that the geometric content of
superconductivity extends beyond the geometry of single-particle Bloch
states. Pairing generates its own emergent geometric structures whose
influence persists from the two-body problem to the collective dynamics
of the many-body state, where the pair effective mass enters a wide
range of physical observables. We therefore identify pair quantum
geometry as an organizing principle for superconductivity beyond
``conventional'' band geometry. More broadly, our work establishes a
hierarchy of geometric concepts, i.e., band geometry, pair geometry, and
Cooper-pair geometry, that govern quantum dynamics at successively
higher levels of organization. We anticipate that related geometric
structures will emerge in a wide range of interacting quantum systems,
including excitonic, photonic, and strongly correlated multiband
platforms.

We also note an intrinsic limitation of the present framework: it
requires a well-defined, momentum-resolved pairing kernel, exact for the
two-body vacuum problem and applicable within the Gaussian-fluctuation
framework near $T_c$, where the fluctuation action is truncated at
quadratic order, and whose inputs $(T_c,\mu)$ rely on the
further approximations of Sec.~\ref{sec: sc}. We do not expect this
description to remain controlled in genuinely strongly correlated
regimes where non-Gaussian fluctuation contributions become important.

The connection between pair geometry and the superfluid
weight has recently been established in Ref.~\cite{kalkan2026} at every
temperature below $T_c$. There, the superfluid weight is set by the
small-momentum curvature of the eigenvalue branch of the static Gaussian
pair-fluctuation kernel that evolves from the Goldstone mode, and the
relaxation correction to the frozen-pairing stiffness is a
stiffness-weighted quantum metric of the corresponding soft eigenvector,
the superconducting-state counterpart of the pair geometry identified
here. Extending the present
effective-mass framework to the low-energy collective modes at zero
temperature, which is the subject of ongoing work, would connect the two
descriptions at the level of the Cooper-pair dynamics.

Finally, we also note a recent study~\cite{sun26} that appeared after 
our initial submission and connects the mode geometry of the static 
fluctuation kernel to paraconductivity and the fluctuational anomalous 
Hall effect within time-dependent Ginzburg-Landau theory.

\begin{acknowledgments}
We acknowledge support from the U.S. Air Force Office of Scientific
Research (AFOSR) under Grant No.~FA8655-24-1-7391.
\end{acknowledgments}

\appendix

\section{Alternative formulation of Eq.~\eqref{eq: pairmass}}
\label{sec:appA}

In this Appendix, we present an alternative, but fully equivalent, 
derivation of Eq.~\eqref{eq: pairmass} based on the on-shell 
differentiation of the secular equation~\eqref{eqn:secularG} and 
the pole condition~\eqref{eqn:poleL}. 
As established in the main text, the two-body bound states are 
governed by the kernel matrix $\mathbb{G}_{1\mathbf{q}}$ defined 
in Eqs.~\eqref{eq:kernel} and~\eqref{eq: G1q}, whose eigenvalues 
$\lambda_{\zeta\mathbf{q}}$ determine the spectral properties of the 
pair manifold as functions of $\mathbf{q}$. Specifically, the bound-state 
energy surface $E = E_{1\mathbf{q}}$ of interest corresponds to 
the trajectory along which exactly one isolated eigenvalue vanishes, 
i.e., the pole condition
$
\lambda_{1\mathbf{q}} = 0,
$
for a given $\mathbf{q}$. All remaining eigenvalues remain finite, 
ensuring that the bound-state branch remains spectrally isolated from 
the other bound states and from the two-particle continuum. 
Consequently, this energy branch can be tracked implicitly through the 
roots of the secular equation
\begin{equation}
I \equiv I(\mathbf{q},E_{1\mathbf{q}})
= \det \mathbb{G}_{1\mathbf{q}}
= \prod_{\zeta}\lambda_{\zeta\mathbf{q}}
= 0.
\label{eq: a1}
\end{equation}
At the bound-state pole, the kernel matrix develops a zero mode,
$
\mathbb{G}_{1\mathbf{q}} |v_{1\mathbf{q}}\rangle = 0,
$
with normalized null eigenvector $|v_{1\mathbf{q}}\rangle$.
Similar to the main text, we assume that the branch extremum is located 
at $\mathbf{q} = \mathbf{0}$ and $E = E_0$.

Since the secular equation
$
I(\mathbf{q},E_{1\mathbf{q}})=0
$
holds identically along the bound-state manifold, its total derivative
with respect to $q_i$ must vanish for all $\mathbf{q}$,
i.e.,
$
\frac{dI}{dq_i}=0,
$
implying
$
\partial_{q_i}I
+
\partial_E I\,\partial_{q_i}E
=0.
$
Assuming that Eq.~\eqref{eq: boundstate} holds for the two-body branch
of interest, the first energy derivative satisfies
$
\partial_{q_i}E=0
$
strictly at the branch extremum, which immediately implies that the first
derivative of the secular equation also vanishes there, i.e.,
$
\partial_{q_i} I = 0
$
at the branch extremum.
Total differentiation of the above identity once more with respect to
$q_j$, which also holds for all $\mathbf{q}$ through the condition
$
\frac{d^2I}{dq_jdq_i}=0,
$
yields the full expression
$
\partial^2_{q_iq_j}I
+
\partial^2_{Eq_i}I\,\partial_{q_j}E
+
\big(
\partial^2_{Eq_j}I
+
\partial^2_EI\,\partial_{q_j}E
\big)\partial_{q_i}E
+
\partial_EI\,\partial^2_{q_iq_j}E
=0.
$
Evaluating this relation strictly at the branch extremum, all cross terms
proportional to
$
\partial_{q_i}E
$
vanish, reducing it to
$
\partial^2_{q_iq_j}I
+
\partial_EI\,\partial^2_{q_iq_j}E
=0.
$
Thus, the matrix elements of the inverse effective mass tensor
$\mathbf{M}_{2b}^{-1}$ are given by
\begin{equation}
\left(M_{2b}^{-1}\right)_{ij}
=
-\frac{\partial^2_{q_iq_j}I}
{\partial_EI},
\label{eq: a2}
\end{equation}
evaluated at the branch extremum. This relation demonstrates that
$\mathbf{M}_{2b}^{-1}$ is fully determined by the local curvature
structure of the secular equation near the bound-state pole.

To establish the equivalence with the adjugate-based formulation derived
in the main text, we first express the secular equation as a product over
the eigenvalues of the kernel matrix, as shown in Eq.~\eqref{eq: a1}.
Differentiating $I$ with respect to energy then yields
$
\partial_E I
=
\mathcal{C}_\mathbf{q}^{2b}\,
\partial_E\lambda_{1\mathbf{q}},
$
where
$
\mathcal{C}_\mathbf{q}^{2b}
=\prod_{\zeta \neq 1}
\lambda_{\zeta\mathbf{q}}
$
denotes the finite and nonzero product of the remaining auxiliary
eigenvalues. Similarly, applying a second-order momentum derivative
to the product representation and evaluating it at the branch extremum
generates a direct curvature contribution
$
\mathcal{C}_\mathbf{0}^{2b}\,
\partial^2_{q_iq_j}\lambda_{1\mathbf{0}},
$
together with additional terms involving products of first-order
derivatives of the vanishing mode,
$
\partial_{q_i}\lambda_{1\mathbf{0}}.
$
Since the pole condition~\eqref{eqn:poleL} holds identically along 
the bound-state manifold, its total derivative with respect to $q_i$ 
must vanish for all $\mathbf{q}$, i.e.,
$
\frac{d\lambda_{1\mathbf{q}}}{dq_i}=0,
$
implying
$
\partial_{q_i}\lambda_{1\mathbf{q}}
+
\partial_E\lambda_{1\mathbf{q}}\,
\partial_{q_i}E
=
0.
$
Thus, assuming that Eq.~\eqref{eq: boundstate} holds for the two-body
branch of interest, we conclude that the first-order derivative vanishes,
$
\partial_{q_i}\lambda_{1\mathbf{0}}=0,
$
at the branch extremum. This condition is also necessary for
$
\partial_{q_i}I=0
$
to hold at the branch extremum.
As a result, all terms containing products of first-order derivatives
drop out, and the second derivative of the determinant reduces to
$
\partial^2_{q_iq_j}I
=
\mathcal{C}_\mathbf{0}^{2b} \,
\partial^2_{q_iq_j}\lambda_{1\mathbf{0}}
$
at the branch extremum. Substituting these spectral derivatives into
Eq.~\eqref{eq: a2} causes the auxiliary factor 
$\mathcal{C}_\mathbf{0}^{2b}$ to cancel identically, yielding
\begin{equation}
\left(M_{2b}^{-1}\right)_{ij}
=
-\frac{\partial^2_{q_iq_j}\lambda_{1\mathbf{0}}}
{\partial_E\lambda_{1\mathbf{0}}},
\label{eq: a3}
\end{equation}
evaluated at the branch extremum. Thus, $\mathbf{M}_{2b}^{-1}$ is
governed entirely by the local curvature structure of the isolated
zero eigenvalue defining the bound-state pole. We also note that an
alternative, but fully equivalent, derivation of Eq.~\eqref{eq: a3}
follows from differentiating the pole condition once more with respect
to $q_j$, and evaluating the result at the branch extremum, i.e., by setting
$
\frac{d^2\lambda_{1\mathbf{0}}}{dq_jdq_i}=0.
$

To explicitly connect this curvature to the underlying quantum geometry
of the auxiliary eigenvectors, we employ the Hellmann-Feynman relation
$
\partial_{q_i}\lambda_{1\mathbf{q}}
=
\langle v_{1\mathbf{q}}|
\partial_{q_i}\mathbb{G}_{1\mathbf{q}}
|v_{1\mathbf{q}}\rangle,
$
and differentiate it once more with respect to $q_j$. This generates
the direct second derivative of the kernel together with two first-order
state-variation terms of the form
$
\langle \partial_{q_j}v_{1\mathbf{q}}|
\partial_{q_i}\mathbb{G}_{1\mathbf{q}}
|v_{1\mathbf{q}}\rangle
$
and its Hermitian conjugate. Invoking the identity
$
\partial_{q_j}\mathbb{G}_{1\mathbf{q}}
|v_{1\mathbf{q}}\rangle
=
-
\mathbb{G}_{1\mathbf{q}}
|\partial_{q_j}v_{1\mathbf{q}}\rangle,
$
these variation terms combine into the real-valued projection
$
-2\operatorname{Re}
\langle \partial_{q_i}v_{1\mathbf{q}}|
\mathbb{G}_{1\mathbf{q}}
|\partial_{q_j}v_{1\mathbf{q}}\rangle.
$
Furthermore, expressing the denominator in Eq.~\eqref{eq: a3} through
the Hellmann-Feynman relation with respect to the energy parameter,
$
\partial_E\lambda_{1\mathbf{q}}
=
\langle v_{1\mathbf{q}}|
\partial_E\mathbb{G}_{1\mathbf{q}}
|v_{1\mathbf{q}}\rangle,
$
we obtain
\begin{equation}
\left(M^{-1}_{2b}\right)_{ij}
=
\frac{
\langle v_{1\mathbf{0}}|
\partial^2_{q_iq_j}\mathbb{G}_{1\mathbf{0}}
|v_{1\mathbf{0}}\rangle
-
2\operatorname{Re}
\langle \partial_{q_i}v_{1\mathbf{0}}|
\mathbb{G}_{1\mathbf{0}}
|\partial_{q_j}v_{1\mathbf{0}}\rangle
}{
\langle v_{1\mathbf{0}}|
\partial_{-E_0}\mathbb{G}_{1\mathbf{0}}
|v_{1\mathbf{0}}\rangle
},
\end{equation}
which is equivalent to Eq.~\eqref{eq: pairmass} upon using
Eq.~\eqref{eq: G1q}.

\section{Conditions for non-trivial pair quantum geometry}
\label{sec:apppair}

The geometric contribution in Eq.~\eqref{eq: pairmass} is governed by the
branch-resolved two-body quantum metric
$
g_{1 \mathbf{0}, 2b}^{\zeta ij}
$
with $\zeta \ne 1$.
To identify the conditions under which this quantity vanishes, it is
useful to decompose the momentum derivative of the null vector into
components parallel and orthogonal to $|v_{1\mathbf{q}}\rangle$:
$
|\partial_{q_i}v_{1\mathbf{q}}\rangle
=
\langle v_{1\mathbf{q}}|\partial_{q_i}v_{1\mathbf{q}}\rangle
|v_{1\mathbf{q}}\rangle
+
|\partial_{q_i}v_{1\mathbf{q}}\rangle_\perp.
$
Since
$
g_{1 \mathbf{0}, 2b}^{\zeta ij}
=
2\,\mathrm{Re}
\big[
\langle\partial_{q_i}v_{1\mathbf{0}}|v_{\zeta\mathbf{0}}\rangle
\langle v_{\zeta\mathbf{0}}|\partial_{q_j}v_{1\mathbf{0}}\rangle
\big]
$
involves projections onto the non-null eigenmodes
$|v_{\zeta\neq1,\mathbf{0}}\rangle$, which are orthogonal to
$|v_{1\mathbf{0}}\rangle$, only the orthogonal component contributes:
$
g_{1 \mathbf{0}, 2b}^{\zeta ij}
=
2\,\mathrm{Re}
\big[
_\perp\langle\partial_{q_i}v_{1\mathbf{0}}|
v_{\zeta\mathbf{0}}\rangle
\langle v_{\zeta\mathbf{0}}
|\partial_{q_j}v_{1\mathbf{0}}\rangle_\perp
\big].
$
Thus, the geometric contribution vanishes whenever
$
|\partial_{q_i}v_{1\mathbf{q}}\rangle_\perp=0.
$

One mechanism that guarantees this condition is a pairing texture that
is independent of momentum up to an overall phase in the
long-wavelength ($\mathbf{q}\to\mathbf{0}$) limit,
$
|v_{1\mathbf{q}}\rangle
=
e^{i\phi_\mathbf{q}}|v_1\rangle,
$
with a momentum-independent $|v_1\rangle$, of which uniform pairing,
$
|v_1\rangle = \frac{1}{\sqrt{N_b}}(1,1,\ldots,1)^\mathrm{T},
$
is a special case. The derivative then reduces to
$
|\partial_{q_i}v_{1\mathbf{q}}\rangle
=
i(\partial_{q_i}\phi_\mathbf{q})|v_{1\mathbf{q}}\rangle,
$
which is purely parallel to $|v_{1\mathbf{q}}\rangle$. Consequently,
$
|\partial_{q_i}v_{1\mathbf{q}}\rangle_\perp=0,
$
and therefore
$
g_{1 \mathbf{0}, 2b}^{\zeta ij}=0
$
for all $\zeta\neq1$. Physically, a center-of-mass momentum boost alters
only the overall phase of the pairing state and leaves its sublattice
structure unchanged. The essential condition is thus the
momentum dependence of the texture, not its non-uniformity at a given
$\mathbf{q}$, since a non-uniform but $\mathbf{q}$-independent texture also
gives a vanishing pair geometry. Since $\beta_{S\mathbf{q}}$ plays the
role of a pairing order parameter, a nontrivial pair geometry is
ultimately rooted in the momentum-dependent sublattice
texture of these amplitudes, which is therefore a necessary
condition. Since a change of embedding multiplies $\beta_{S\mathbf{q}}$ by
$e^{-i\mathbf{q}\cdot\mathbf{d}_S}$, this condition refers to the chosen
embedding. Uniform pairing in one embedding becomes the
$\mathbf{q}$-dependent texture $e^{-i\mathbf{q}\cdot\mathbf{d}_S}/\sqrt{N_b}$
in another, which is how the two terms of Eq.~\eqref{eq: pairmass}
redistribute while their sum is invariant (Sec.~\ref{sec:tbp}). It is
not, however, sufficient. Even when the pairing texture is strongly
momentum dependent, symmetry constraints may force the
geometric contribution to vanish after summing over all intermediate
channels. For example, we verified numerically that the perovskite
lattice exhibits precisely this behavior~\cite{weeks10}.
A second, independent mechanism follows from inversion symmetry of the
pair kernel $\mathbb{G}_{1\mathbf{q}}$ in the long-wavelength limit. 
If
$
\mathbb{G}_{1\mathbf{q}}
=
\mathbb{G}_{1,-\mathbf{q}},
$
then
$
\partial_{q_i}\mathbb{G}_{1\mathbf{0}}=0
$
at the branch extremum. Using
$
\mathbb{G}_{1\mathbf{q}}|\partial_{q_i}v_{1\mathbf{q}}\rangle
=
-\partial_{q_i}\mathbb{G}_{1\mathbf{q}}|v_{1\mathbf{q}}\rangle,
$
one finds at $\mathbf{q}=\mathbf{0}$ that
$
|\partial_{q_i}v_{1\mathbf{0}}\rangle
$
has no projection onto the non-null sector of
$
\mathbb{G}_{1\mathbf{0}}.
$
Its overlap with every non-null eigenmode therefore vanishes, implying
$
g_{1 \mathbf{0}, 2b}^{\zeta ij}=0
$
for all $\zeta\neq1$.

In summary, the pair-geometry contribution vanishes whenever the pairing
texture is momentum independent up to a phase or the pair
kernel is inversion symmetric in the long-wavelength limit. Therefore,
a momentum-dependent pairing texture and inversion-symmetry
breaking in the pair sector are both necessary conditions for a finite
geometric contribution to Eq.~\eqref{eq: pairmass}, although neither is
generally sufficient on its own.

\section{Alternative formulation of Eq.~\eqref{eq: Cooperinversemass}}
\label{sec:appB}

In this Appendix, we present an alternative but fully equivalent
derivation of Eq.~\eqref{eq: Cooperinversemass}, based on the on-shell
differentiation of the secular equation~\eqref{eqn:secularK} and the
pole condition~\eqref{eqn:poleC}, rather than Jacobi's formula
Eq.~\eqref{eqn:jacobi}. Since the derivation closely parallels that of
App.~\ref{sec:appA}, with the Hermitian kernel $\mathbb{G}_{1\mathbf{q}}$
replaced by the non-Hermitian kernel $\mathbb{K}_{1\mathbf{q}}$ and the
corresponding Hermitian formalism replaced by its biorthogonal analogue,
we outline only the key steps. The secular equation
Eq.~\eqref{eqn:secularK} holds identically along the collective-mode
dispersion. Applying the same on-shell differentiation procedure used in
App.~\ref{sec:appA} to
$
J \equiv J(\mathbf{q}, z_{1\mathbf{q}})
= \det \mathbb{K}_{1\mathbf{q}} = 0,
$
which is analytic in $z_{1\mathbf{q}}$ on the second sheet,
and using $\partial_{q_i}z_{1\mathbf{0}} = 0$ at $\mathbf{q}=\mathbf{0}$,
gives
\begin{equation}
(M_{\mathrm{Cp}}^{-1})_{ij} =
-\frac{\partial^2_{q_iq_j} J}{\partial_z J}.
\label{eq: b1}
\end{equation}
Writing
$
J = \chi_{1\mathbf{q}} \mathcal{C}^{\mathrm{Cp}}_{\mathbf{q}},
$
and noting that $\partial_{q_i}\chi_{1\mathbf{0}} = 0$ at the mode
extremum, which follows from the collective-mode pole condition in the
same way that
$
\partial_{q_i}\lambda_{1\mathbf{0}} = 0
$
follows from the bound-state pole condition discussed in
App.~\ref{sec:appA}, the spectral factor
$\mathcal{C}^{\mathrm{Cp}}_{\mathbf{0}}$ cancels identically, reducing
Eq.~\eqref{eq: b1} to
\begin{equation}
(M^{-1}_{\mathrm{Cp}})_{ij}
=
-\frac{\partial^2_{q_i q_j} \chi_{1\mathbf{0}}}
{\partial_{z} \chi_{1\mathbf{0}}}.
\label{eq: b2}
\end{equation}
Applying the biorthogonal Hellmann-Feynman theorem,
$\partial_{q_i}\chi_{1\mathbf{q}}
=
\langle u^L_{1\mathbf{q}}|
\partial_{q_i}\mathbb{K}_{1\mathbf{q}}
|u^R_{1\mathbf{q}}\rangle,
$
differentiating once more with respect to $q_j$, and expressing the
eigenvector derivatives through the null-space identities
$
\mathbb{K}_{1\mathbf{0}}|\partial_{q_i}u^R_{1\mathbf{0}}\rangle
=
-\partial_{q_i}\mathbb{K}_{1\mathbf{0}}|u^R_{1\mathbf{0}}\rangle
$
and
$
\langle\partial_{q_i}u^L_{1\mathbf{0}}|\mathbb{K}_{1\mathbf{0}}
=
-\langle u^L_{1\mathbf{0}}|\partial_{q_i}\mathbb{K}_{1\mathbf{0}},
$
followed by projection onto the non-null sector
$
{\mathbb{K}_{1\mathbf{0}}
=
\sum_{\eta \neq 1} \chi_{\eta\mathbf{0}}
|u^R_{\eta\mathbf{0}}\rangle\langle u^L_{\eta\mathbf{0}}|},
$
yields
$
\partial^2_{q_i q_j} \chi_{1\mathbf{0}}
=
\langle u^L_{1\mathbf{0}}|
\partial^2_{q_i q_j}\mathbb{K}_{1\mathbf{0}}
|u^R_{1\mathbf{0}}\rangle
-
\sum_{\eta \neq 1} \chi_{\eta\mathbf{0}}
\big(
\langle\partial_{q_i}u^L_{1\mathbf{0}}|u^R_{\eta\mathbf{0}}\rangle
\langle u^L_{\eta\mathbf{0}}|\partial_{q_j}u^R_{1\mathbf{0}}\rangle
+
\langle\partial_{q_j}u^L_{1\mathbf{0}}|u^R_{\eta\mathbf{0}}\rangle
\langle u^L_{\eta\mathbf{0}}|\partial_{q_i}u^R_{1\mathbf{0}}\rangle
\big)
$
at $\mathbf{q}=\mathbf{0}$. Substituting this result, together with
$
\partial_z \chi_{1\mathbf{0}}
=
\langle u^L_{1\mathbf{0}}|
\partial_z \mathbb{K}_{1\mathbf{0}}
|u^R_{1\mathbf{0}}\rangle,
$
into Eq.~\eqref{eq: b2} reproduces Eq.~\eqref{eq: Cooperinversemass}
exactly.

\bibliography{refs}

\end{document}